\documentclass[aps,prl,twocolumn]{revtex4-1}
\usepackage{amsmath,soul}
\usepackage{amssymb}
\usepackage{graphicx}
\usepackage{epstopdf}
\usepackage[colorlinks=true]{hyperref}

\usepackage{physics}
\usepackage{mathrsfs}
\usepackage{comment}
\usepackage{bm}

\newcommand{\sectionn}[1]{{\textit{#1}}}
\renewcommand\({\begin{equation}}	
\renewcommand\){\end{equation}}

\renewcommand\[{\begin{eqnarray}}	
\renewcommand\]{\end{eqnarray}}

\newcommand{\bb}[1]{{\pmb{\left[\vphantom{#1}\right.} #1 \pmb{\left.\vphantom{#1}\right]}}}
\newcommand{\bunderline}[1]{\underline{#1\mkern-4mu}\mkern4mu }

\newcommand{\al}[1]{\begin{aligned}#1\end{aligned}}
\newcommand{\eq}[1]{\begin{equation}#1\end{equation}}

\begin{document}

\title{Bell-state generation for spin qubits via  dissipative coupling  }

\author{Ji Zou}
\email{jzeeb@ucla.edu}
\author{Shu Zhang}
\author{Yaroslav Tserkovnyak}
\affiliation{Department of Physics and Astronomy, University of California, Los Angeles, California 90095, USA}

\begin{abstract}
We theoretically investigate the dynamics of two
spin qubits
interacting with a magnetic medium.
A systematic
formal framework for this qubit-magnet hybrid system  is developed in terms of 
the steady-state properties of the magnetic medium.
Focusing on the induced dissipative coupling between the spin qubits, we show how a sizable long-lived entanglement can be established via the magnetic environment, in the absence of any coherent coupling.
Moreover, we demonstrate that maximally-entangled two-qubit states (Bell states) can be 
achieved
in this scheme when complemented
by proper postselection. 
In this situation, the  time evolution of the entanglement is governed by a non-Hermitian Hamiltonian, where dynamical phases are separated by an exceptional point.
The resultant Bell state
is robust against weak random perturbations and 
does not require the preparation of 
a particular initial state. 
Our study may find applications in quantum information science, quantum spintronics, and for sensing of nonlocal quantum correlations.
\end{abstract}

\date{\today}
\maketitle

\underline{\sectionn{Introduction.}}|Entanglement between individually addressable qubits is the key to many quantum-information processes~\cite{nielsen, bellrmp}.
The realization of qubits 
has been achieved in several systems, such as  
trapped atoms~\cite{Blinov2004nature, volz2006prl, Blatt2008nature, Bruzewicz2019aip}, 
quantum dots~\cite{PhysRevA.57.120, Basso2019prl, Qiao:2020ncom}, superconducting circuits~\cite{Wendin_2017}, and nitrogen-vacancy (NV) centers \cite{NV2013}, etc.
For example, the  NV qubit has a long coherence time and
a good performance in the initialization and readout of spin states~\cite{chu2015nv,Bar-Gill:2013ur,Balasubramanian:2009ub}.
However, because the direct dipolar interactions between NVs extend only up to tens of nanometers, the generation of entanglement between distant qubits has been one of the  main adversities in building a scalable platform for practical applications.
A potential solution to this problem is to exploit hybrid quantum devices~\cite{Wallquist_2009}, where qubits are interfaced with a solid-state system~\cite{chunhui2020,Du195,Andrich:2017vq,Sar:2015tr,wolfe2014,wolfe2016apl}.
The latter, being long-range correlated, can act as a medium to induce an effective coherent coupling between the qubits, based on which certain two-qubit gates can be implemented~\cite{Daniel2012prx,Daniel2013prx}. 
Meanwhile, the presence of a medium also enhances dissipation effects. To achieve a finite entanglement between qubits, the timescale set by the coherent coupling needs to be shorter than that of the local qubit relaxation.
The competition between the two has thus been the focus of recent investigations~\cite{Daniel2012prx,Daniel2013prx,contrerasprb2008,PhysRevLett.121.187204,PhysRevB.99.140403,guidoprb2019,Candido_2020,derekprl2020,Fukamiarxiv2021}.

Dissipation, however, is not always detrimental to quantum effects. Entanglement generation in an open quantum system by environment engineering was first discussed in the context of  quantum optics~\cite{Zollerprl1996, Pleniopra1999}. It was shown formally that two qubits can be entangled by undergoing Markovian dissipative dynamics~\cite{BenattiPRL2003}. Various proposals have been made to realize this, mainly in quantum optical and electronic systems~\cite{Diehl2008natphy, Lin2013nature, kim2016prl, Kordas2012EPL, Dico2004pra, yokoshi2017, Kienzler2015science, cirac2011prl, benito2016prb, Li_pra_2012, Wang_pra_2020, prr_ullah_2022}.
In addition, dissipation is also investigated as a resource for  quantum error correction~\cite{Reiter2017naturecom, kapit2016prl, cohen2014pra, freeman2017pra, Leghtas2015science, Shankar2013nature} and other quantum information tasks~\cite{Mirrahimi_2014, modi2011prx,modi2018njp, botzungprb2021}. Non-Hermitian Hamiltonians, frequently invoked to handle dissipative effects in the Hamiltonian form, can exhibit exceptional points~\cite{Bender_2007} that have been shown to be sweet spots to enhance entanglement~\cite{lin2016pra, yuan2020prl}.

\begin{figure}
  \includegraphics[width = \linewidth]{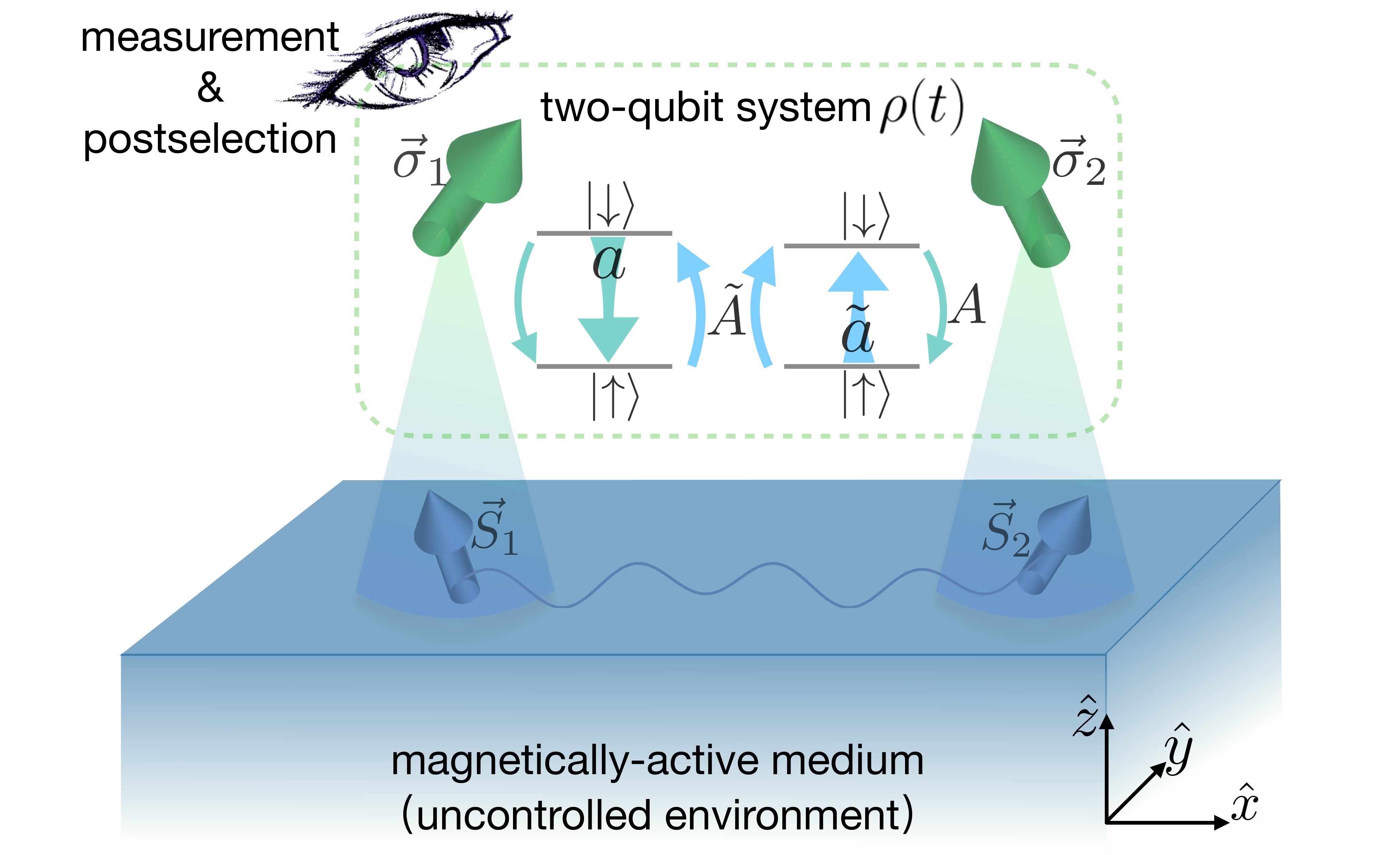}
   \caption{A system composed of two spin qubits is coupled with a magnetic environment, which induces local relaxations $a$, $\tilde{a}$,
   mediates dissipative couplings $A$, $\tilde{A}$, as well as coherent couplings between two qubits.  
    The two qubits may achieve a stable entangled state with large enough $A$ and $\tilde{A}$, and even a Bell state with the help of measurement and postselection.
}
   \label{fig1}
\end{figure} 

In this work, we discuss the dissipative coupling and entanglement generation induced by a generic noisy magnetic medium, in a hybrid quantum system sketched in Fig.~\ref{fig1}. In particular, we demonstrate that, when complemented by proper postselections, a Bell state can be generated through an exceptional point in the time evolution governed by a non-Hermitian Hamiltonian.
The qubits can be NVs or other isolated quantum defects and the medium is a generic solid-state system emitting magnetic field noise, which can arise from fluctuations of spin or pseudospin degrees of freedom~\cite{suzy2021}.
Since many magnetic materials with different correlation properties are generally available, artificial design of the environment is not required as a first step, while spintronic engineering and tunability are promising for future studies. 

To treat the induced coherent and dissipative couplings in a unified manner, we derive the full master equation~\cite{Heinz, Lidar2019} that determines the 
time evolution of the qubit entanglement.
Specifically, two distinct types of dissipation are identified, bearing analogy to the local damping and the spin pumping-mediated viscosity in the classical spin dynamics~\cite{epyaroslav}:
One is the local relaxation, which originates in energy and information exchanges 
between a single qubit and the medium. The other is the dissipative coupling between the qubits induced by the correlated medium they both couple to.
While the former is detrimental to quantum entanglement, we show the latter can help to establish a steady entanglement between qubits, even in a pessimistic scenario where the coherent coupling is absent.   
The long-time behavior of the qubits reflects a phase transition, as a function of system parameters.
When the dissipative coupling is comparable to the local relaxation, the Lindbladian evolution induced by the medium can result in sizable robust entanglement between the qubits. This can be achieved for qubit separation on a lengthscale dictated by the relevant excitations responsible for dissipation (such as magnons for a magnetically ordered medium).


\underline{\sectionn{Model.}}|Let us consider an illustrative model consisting of two spin qubits weakly coupled to a magnet, with the following Hamiltonian:
\( H=H_{\text{S}}+H_{\text{E}}+H_{\text{SE}}.\label{totalH} \)
Here, $H_{\text{S}}=-( \Delta_1\sigma_1^z+\Delta_2\sigma_2^z)/2$ is the Hamiltonian for the system with two qubits subjected to magnetic fields $\Delta_1$ and $\Delta_2$, respectively, along the $z$ direction, $H_{\text{E}}$ is an unspecified  Hamiltonian of the medium as an environment for the system, and $H_{\text{SE}}=\lambda (\Vec{\sigma}_1\cdot \Vec{S}_1  + \Vec{\sigma}_2\cdot \Vec{S}_2 )$ 
describes the system-environment interaction with coupling strength $\lambda$, where $ \vec{\sigma}_i$ stands for the Pauli matrices of the $i$th qubit, and $\vec{S}_i$ for local spin density operators it couples to within the medium.
Without loss of generality,  we assume $\Delta_1\geq \Delta_2\geq 0$. We will consider an axially-symmetric environment $H_{\text{E}}$ in spin space, while a generalization would be straightforward. It would also be straightforward to generalize the treatment to the dipolar coupling between the qubit and the medium \cite{PhysRevLett.121.187204,PhysRevB.99.140403}.

The following Lindblad master equation of the density matrix of the two-qubit system can be derived microscopically based on the Born and Markov approximations: 
\(  \dv{}{t}\rho = -i\big[H_{\text{S}}+H_{\text{eff}} ,\rho \big]  - \mathcal{L}[\rho].  \label{master} \)
Leaving the derivation to the supplemental materials~\cite{entsm}, 
we start with a phenomenological understanding of it on symmetry grounds.
Here, $H_{\text{eff}}$ is the medium-induced effective coherent coupling between qubits, participating in the unitary system evolution, while $\mathcal{L}[\rho]$ is the  dissipative Lindbladian expanded in the usual form:
 \(  \mathcal{L}[\rho]= \sum_{nm}h_{nm}  \big(  \mathcal{O}^\dagger_m \mathcal{O}_n \rho + \rho \mathcal{O}^\dagger_m \mathcal{O}_n -2 \mathcal{O}_n \rho \mathcal{O}^\dagger_m   \big), \label{L} \) 
where the coefficient matrix $h$ is Hermitian and positive-semidefinite~\cite{Heinz, Lidar2019}, and  $\mathcal{O}\!\! =\!\! ( \sigma_1^- , \sigma_2^- , \sigma_1^+ , \sigma_2^+ , \sigma_1^z , \sigma_2^z )$ comprises qubit operators.

The most general form of $H_{\text{eff}}$, allowed by the axial symmetry, is $H_{\text{eff}}=\mathcal{J}_z\sigma^z_1\sigma^z_2+\mathcal{J}_\perp (\sigma_1^x\sigma_2^x  + \sigma_1^y\sigma_2^y)+D \hat{z}\cdot \vec{\sigma}_1\times \vec{\sigma}_2$,  a summation of an XXZ model and a Dzyaloshinskii-Moriya (DM) interaction term~\cite{entsm}.  The DM interaction must vanish if, for example, the structure is invariant under $\pi$ $z$-rotation (see Fig.~\ref{fig1} for the coordinate frame). These coherent couplings induced by the magnetic medium can build up a finite entanglement within the timescale inversely proportional to the coupling strength~\cite{entsm}, if it is shorter than the timescale set by dissipation.  
In the limiting case of a full isotropicity in spin space and $\Delta_i= 0$, $H_{\text{eff}}$  is further reduced to a Heisenberg form $H_{\text{eff}} = \mathcal{J}\vec{\sigma}_1\cdot \vec{\sigma}_2$ resembling the RKKY coupling~\cite{isotropicspin}. 
These effective coupling parameters are all real constants determined by the Green's functions of the medium~\cite{entsm}, as is consistent with previous results from  Schrieffer-Wolff transformation~\cite{Daniel2012prx,Daniel2013prx,PhysRevLett.121.187204,PhysRevB.99.140403,guidoprb2019,Candido_2020,derekprl2020,Fukamiarxiv2021}. 
Direct dipolar interaction between qubits is typically negligible, except for very small spacings.

In the dissipative Lindblad part, $h$ is block diagonal due to the axial symmetry. 
In general terms, we have 
\(  h=\mqty(\tilde{a} & \tilde{A} \\ \tilde{A}^* & \tilde{a}) \oplus  \mqty(a & A^* \\ A & a) \oplus  \mqty(d & \mathfrak{D}\\  \mathfrak{D}^* &d),  \)
where $ \tilde{a}, a, d$ and $ \tilde{A}, A,  \mathfrak{D} $ are real and complex  phenomenological parameters, respectively. These parameters represent three types of dissipative effects:
$a$ and $\tilde{a}$ are associated with local decay and the reverse process. They govern local relaxation of individual qubits, giving rise to the relaxation time $T_1$ and contribute to the decoherence time $T_2$ of a single qubit~\cite{entsm}. In contrast, $A$ and $\tilde{A}$ are related to cooperative decay and the reverse process involving both qubits, and are referred to as dissipative couplings, which depend on the distance between the two qubits. They are the focus of this work. $d$ and $\mathfrak{D}$ are pure-dephasing parameters, originating from those terms in $H_{\text{SE}}$ that commute with $H_{\text{S}}$, namely $\lambda\sum_{i=1,2}\sigma_i^zS^z_i$. They only cause information but not energy exchange between the system and the medium, and in practice
may be mitigated by dynamic decoupling~\cite{violaprl1999,khodprl2005,loprb2014,Paz_Silva_2016}.
We neglect pure-dephasing effects in the following discussion, though they may also lead to entanglement between multiple qubits as shown recently~\cite{aash_prl_2022}. 
The  Lindbladian~(\ref{L}) can then be brought into a diagonal form with four quantum-jump operators~\cite{dalibard1992prl, entphase}
\[  \label{eq:jump-operator}
J_1 &= \sqrt{\cfrac{\tilde{a}+|\tilde{A}|}{2}}    \left( \sigma_1^-+\sigma_2^-  \right), \; \; J_2=  \sqrt{\cfrac{\tilde{a}-|\tilde{A}|}{2}}     \left( \sigma_1^--\sigma_2^-  \right),  \nonumber \\
    J_3 &= \sqrt{\cfrac{a+|A|}{2}}  \left( \sigma_1^++\sigma_2^+  \right),  \; \; J_4=    \sqrt{\cfrac{a-|A|}{2}}   \left( \sigma_1^+-\sigma_2^+  \right),   \nonumber
\\ \]
yielding
 \( \underline{\mathcal{L}}[\rho] =   \sum_{i=1}^4 \mathcal{D}_{J_i}[\rho], \)
where the dissipator is defined as $\mathcal{D}_J[\rho]\equiv J^\dagger J \rho + \rho J^\dagger J -2J \rho J^\dagger$.
    
Microscopically, all parameters are given by the Green's functions of the medium in equilibrium~\cite{entsm}, such that the fluctuation-dissipation theorem dictates that they are not independent: $\tilde{a} = e^{-\beta \Delta} a$ and $\tilde{A} = e^{-\beta \Delta} A$, where $\beta = 1/k_B T$ and $\Delta \equiv (\Delta_1+\Delta_2)/2$.
The zero temperature therefore corresponds to $\tilde{a}=\tilde{A}=0$, where only the decay processes survive.
Also, the thermodynamic stability of the magnetic medium imposes
 $   a \geq |A|$ and $\tilde{a} \geq |\tilde{A}| $~\cite{entsm},
which ensures the matrix $h$ is positive-semidefinite.

\underline{\sectionn{Dissipative coupling vs local relaxation.}}|Let us now  explore the entanglement evolution of  two qubits focusing on  
the dissipative effects, by setting ourselves
in a pessimistic situation
where the induced coherent dynamics is absent:
\( \dv{}{t}\rho = -i[ H_{\text{S}},\rho ]- \underline{\mathcal{L}}[\rho]. \label{ma}  \)
Here, we
treat the scenario of  zero temperature $\tilde{a}=\tilde{A}=0$  
analytically
to demonstrate the effects of local relaxation and dissipative couplings. Numerical results for finite temperature are presented in the Supplemental Material~\cite{entsm}, which do not qualitatively change our conclusion below. 

  \begin{figure}[t]
  \includegraphics[scale=.21]{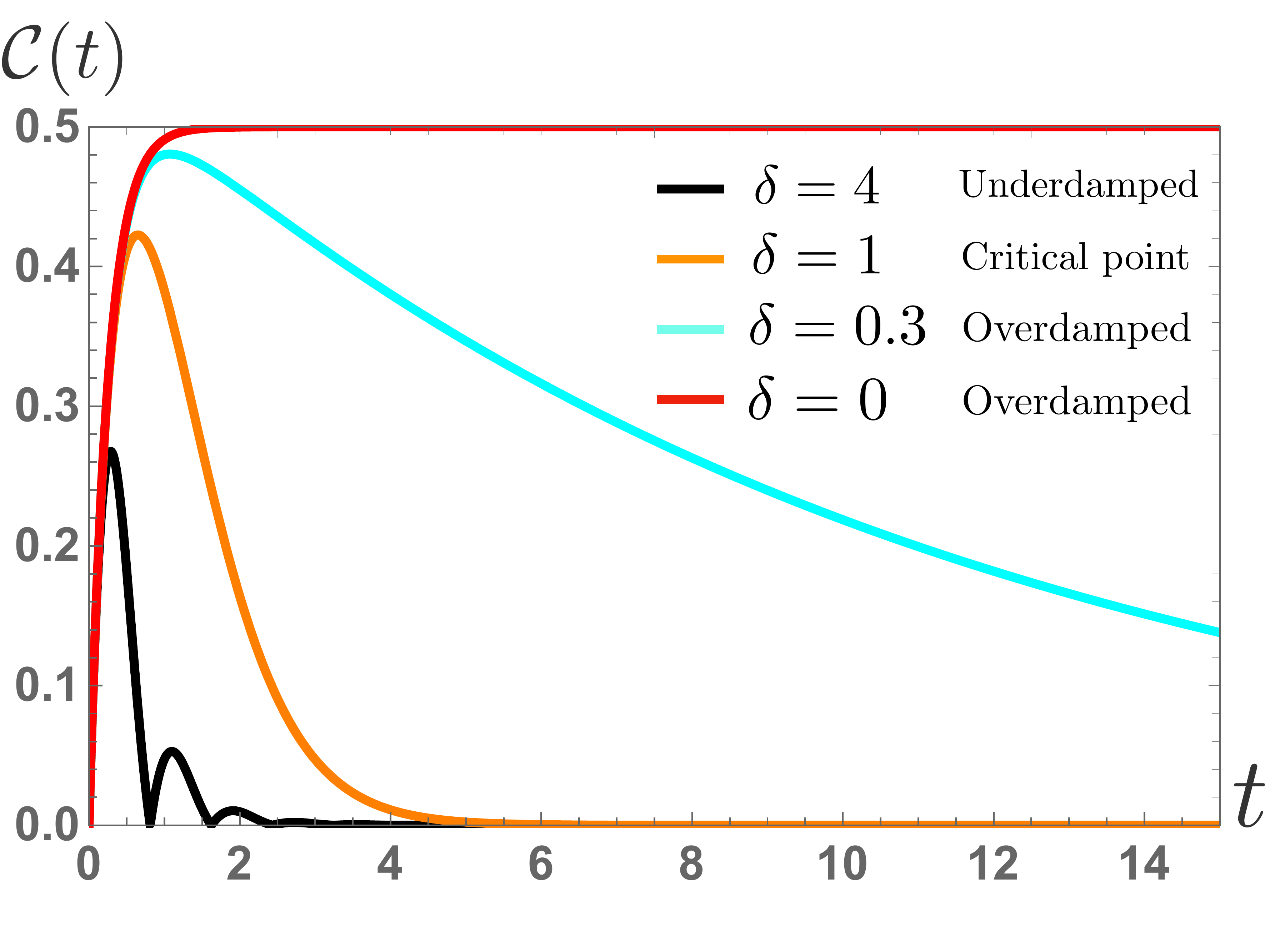}
   \caption{Concurrence of two qubits as a function of time, with initial state $\ket{\uparrow\downarrow}$, where we set both  local dissipation $a$ and dissipative coupling $|A|$  to be 1.  The black curve corresponds to the underdamped quantum regime. The orange curve is  at the critical point $\delta=1$, where entanglement  decays  as $\mathcal{C}(t)\propto te^{-2at}$.  The cyan, $\delta=0.3$, and the red, $\delta=0$, curves are in overdamped quantum regime, where the lifetime of entanglement is extended dramatically. } 
   \label{fig2}
\end{figure}

The qubits are initialized into
a trivial product state, 
taking the example of
$\ket{\uparrow \downarrow}$ for the sake of concreteness. 
We show the master equation~(\ref{ma}) can be reduced to an equation for $x\equiv \Re \bra{\uparrow\downarrow}\rho\ket{\downarrow\uparrow}$~\cite{entsm}:
\( \ddot{x} + 4a\dot{x}+4 \left(\delta^2+a^2-|A|^2 \right)x=0, \) 
where $\delta\equiv (\Delta_1-\Delta_2)/2$ is the  local field asymmetry. This equation resembles a damped oscillator with 
complex characteristic frequencies 
\(  \omega_\pm =\pm 2\omega_0 - i 2a,   \)
where $\omega_0\equiv\sqrt{\delta^2-|A|^2}$. The real part gives the coherent beating of the density matrix elements, while the imaginary part reflects decoherence.
The contribution from local relaxation $-i2a$
leads to a decaying envelope factor $e^{-2at}$ in the entanglement between two qubits (as detailed below), indicating its detrimental effect on quantum coherence as expected. 

We identify three distinct parameter regimes for the quantum dynamics. 
In the underdamped regime,  $\delta>|A|$, 
$\omega_0$ is real valued.
To quantify the time evolution of the entanglement between the two qubits, we calculate the
\textit{concurrence}~\cite{PhysRevLett.80.2245, entsm}
as a function of time:
$\mathcal{C}(t)=2  e^{-2at}   |A\sin\omega_0 t |    \sqrt{\delta^2- |A\cos\omega_0t|^2}/ \omega_0^2 $. See Fig.~\ref{fig2}.
The entanglement oscillates with frequency $2\omega_0$ as the system  decays rapidly to the ground state $\ket{\uparrow\uparrow}$
on the time scale $\tau=1/2a$.  
At the critical point $\delta=|A|$, $\omega_0 = 0$, there is no oscillation.
The concurrence evolves 
as $\mathcal{C}(t)\propto te^{-2a t}$, 
where the final steady state is also $\ket{\uparrow\uparrow}$. As shown in Fig.~\ref{fig2}, we have a larger transient entanglement and the decay process is slowed down moderately compared with the underdamped regime.
  
In the overdamped regime, $\delta<|A|$,  
$\omega_0$ becomes purely imaginary and $\omega_\pm =-2i(a\pm \kappa_0)$, with $\kappa_0=\sqrt{|A|^2-\delta^2}$.     
The time-dependent concurrence is $\mathcal{C}(t)=2 e^{-2at}   |A|\sinh\kappa_0 t  \sqrt{|A\cosh\kappa_0 t|^2 -\delta^2}/\kappa_0^2$.
In addition to a  larger  transient  entanglement, the decay process has been slowed down dramatically. 
On a long time scale $t\gg 1/\kappa_0$, $\mathcal{C}(t) \propto e^{-2(a-\kappa_0)t}$.
The entanglement can last for $\tau=1/2(a-\kappa_0)$, which becomes $\tau=1/2(a-|A|)$ when the two local fields are the same, $\delta=0$.
See Fig.~\ref{fig2}. 
It is clear from this expression of the lifetime $\tau$ that the dissipative coupling $A$ and the local relaxation $a$,
though both originating from the qubits-magnet coupling,
have opposite effects on the quantum entanglement in the nonunitary evolution. 
The local dissipation tends to destroy any quantum coherence whereas the dissipative coupling 
can be exploited to
extend the lifetime of entanglement and 
even realize
steady entangled states. 
With equal local fields $\delta=0$, a finite entanglement can persist for a long time before eventually decaying to zero in the large dissipative coupling regime $|A| \lesssim a$. Based on their (greater) Green's function expressions~\cite{greater, entsm} $ 2 a\! = \!  i \lambda^2 G^>_{S^+_1S^-_1}(\Delta) =i \lambda^2 G^>_{S^+_2S^-_2}(\Delta)$, $2A = i \lambda^2  G^>_{S^+_1S^-_2}(\Delta)$, $|A| \lesssim a$
physically corresponds to the scenario with two qubits placed within a lengthscale dictated by the relevant excitations responsible for dissipation. For example, for qubits coupled to a magnetically ordered medium via processes of magnon absorption and emission, this lengthscale is set by the wavelength of the magnon at frequency $\Delta$.
Furthermore, the concurrence lifetime extends to infinity $\tau\rightarrow \infty$ when $|A|$ reaches its maximal allowed value $|A| = a$,  and thus a steady entanglement is achieved, $\mathcal{C}(\infty)=1/2$ for the final state  
  \(   \rho_{\infty}=\left( \ket{\uparrow\uparrow}\bra{\uparrow\uparrow} + \ket{00}\bra{00}\right)/2.      \)
  Noting that the singlet $\ket{00}$ cannot be evolved to a different state by the operative jump operators or the system Hamiltonian, it is a dark state---the system would stay in this pure state indefinitely. For this reason, a steady-state entanglement can also be reached at finite temperatures
when $a=|A|$, $\tilde{a}=|\tilde{A}|$~\cite{entsm}, 
though with a smaller concurrence. We also remark that finite steady-state entanglement can be achieved in this optimal situation, irrespective of the initial two-qubit state as long as it is not a symmetric state~\cite{entsm}.


We stress that 
the above  critical point and the associated  transition from underdamped to overdamped regime
 result from the dissipative couplings, which
are the main focus of this article. 
We 
next show how to generate Bell states by exploiting this dissipative coupling, when combined with proper measurements and  postselections.

 \begin{figure}[t]
  \includegraphics[scale=.22]{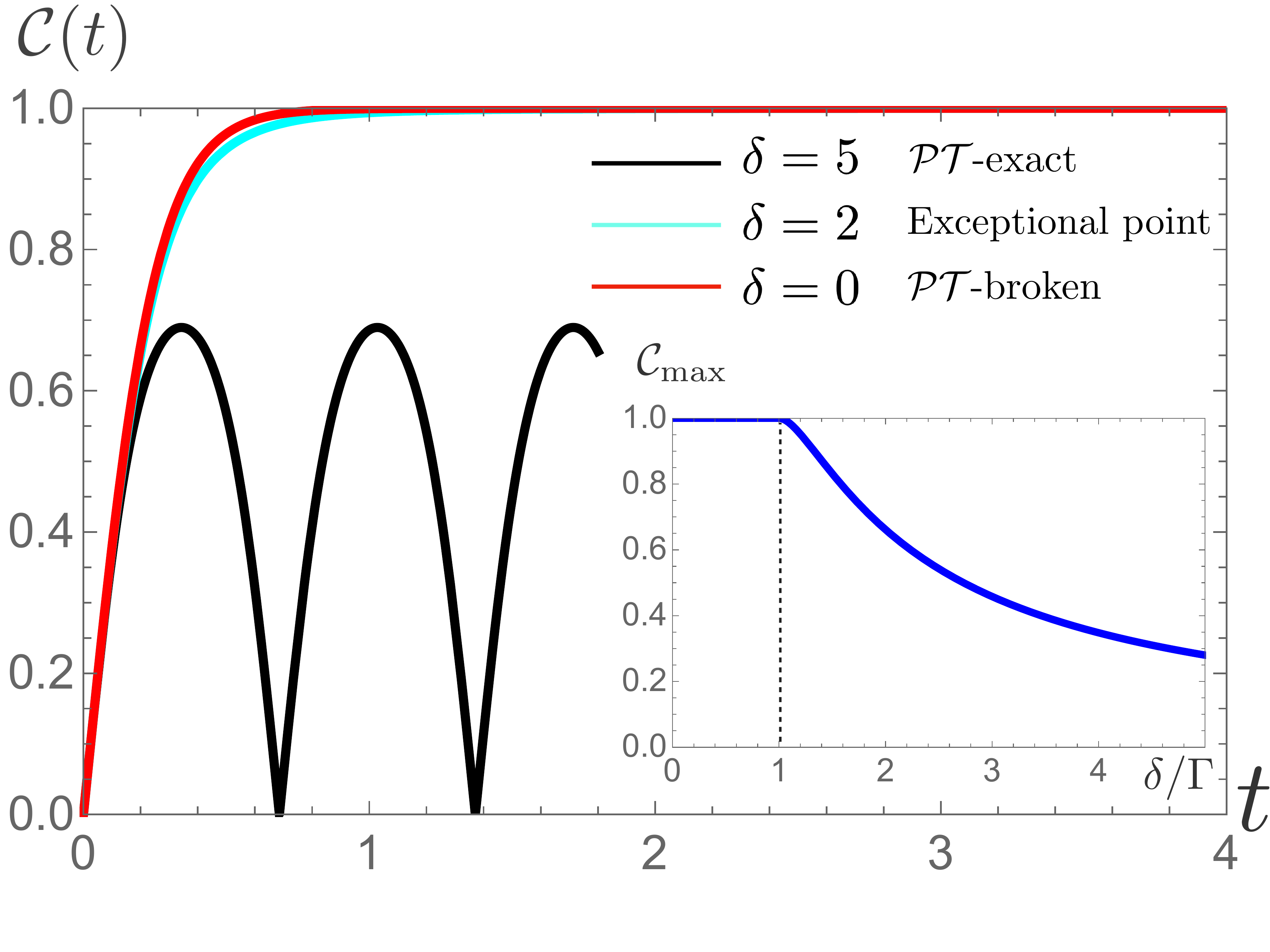}
   \caption{Concurrence of qubits as a function of time with initial state $\ket{\uparrow\downarrow}$ under continuous measurements and postselections. We set $\Gamma=2$. The black curve $\delta=5$ is in the $\mathcal{PT}$-exact regime, where entanglement oscillates and its maximal value is less than 1. At the exceptional point (cyan curve), there is no oscillation and its maximal value is 1. In $\mathcal{PT}$-broken regime (red curve), entanglement is $\mathcal{C}(t)=\tanh 2\Gamma t$. The inset shows the maximal concurrence as a function of $\delta/\Gamma$. } 
   \label{fig4}
\end{figure}

\underline{\sectionn{Non-Hermitian Hamiltonian scheme.}}|Let us turn to the evolution of qubits under measurements (see Fig.~\ref{fig1}), which is often invoked to perform feedback and conditional control as a valuable resource in controlling open quantum systems~\cite{qumeasure,hume2007prl,Negnevitsky2018nature,Jiang2009sci,xin2019prl}. To this end, we rewrite the master equation~(\ref{ma}) in the following form:
\( \dv{}{t}\rho=-i \bb{ \bunderline{H}_{\text{eff}},\rho } +2\sum_{i=1}^4 J_i\rho J_i^\dagger,  \label{drhodt} \)
where \(\bunderline{H}_{\text{eff}}=H_S-i \sum_{i}J_i^\dagger J_i, \label{noneff} \) is a non-Hermitian Hamiltonian.  Correspondingly, the commutator should now be understood as $\bb{\bunderline{H}_{\text{eff}},\rho } \equiv \bunderline{H}_{\text{eff}}\rho- \rho \bunderline{H}_{\text{eff}}^\dagger=[H_S,\rho]-i \{ \sum_i J^\dagger_i J_i,\rho \}$. 
By subjecting the two qubits to continuous measurements of the absolute value of their total spin $z$ component $\underline{\sigma}^z=\sigma_1^z+\sigma_2^z$  and subsequently conditioning the postselection on zero outcomes, we can effectively 
forbid  all quantum jump processes~(\ref{eq:jump-operator}),
as $[J_i,\underline{\sigma}^z]\neq 0$. This monitored dynamics of the two-qubit system formally eliminates the last term in Eq.~\eqref{drhodt} and reduces quantum dynamics to a non-Hermitian Hamiltonian form in the $\underline{\sigma}^z=0$ subspace \cite{Bender_2007}, 
$ d\rho/dt=-i \bb{ \bunderline{H}_{\text{eff}},\rho }$, whose integration is appropriately normalized to give \cite{brody2012prl}: \(\rho(t)=   \frac{ e^{-i \bunderline{H}_{\text{eff}} t} \rho_0 e^{i \bunderline{H}_{\text{eff}}^\dagger t} }{\tr\big(  e^{-i \bunderline{H}_{\text{eff}} t} \rho_0 e^{i \bunderline{H}_{\text{eff}}^\dagger t} \big)}, \label{rhosol}  \)
in terms of the initial qubits state $\rho_0$. 
Since the effective Hamiltonian~\eqref{noneff} conserves the quantum number $\underline{\sigma}^z$, 
$[\bunderline{H}_{\text{eff}}, \underline{\sigma}^z]=0$,
the subspace $\mathcal{H}$ spanned by $\{ \ket{\uparrow\downarrow},\ket{\downarrow\uparrow}\}$
is closed under time evolution.
 The eigenkets of $\bunderline{H}_{\text{eff}}$ in $\mathcal{H}$ are
\[   \ket{s}&=& e^{i\varphi}\sin\theta \ket{\uparrow\downarrow}+\cos\theta \ket{\downarrow\uparrow}, \nonumber \\
         \ket{a}&=&-\cos\theta \ket{\uparrow\downarrow}+  e^{i\varphi}\sin\theta\ket{\downarrow\uparrow},\]
with associated eigenenergies $E_{s(a)}=\mp\omega-i(a+\tilde{a})$.
Here, $\theta$ and $\varphi$  are determined by the sum of dissipative couplings $\Gamma=|A|+|\tilde{A}|$ and  the local field asymmetry $\delta$: For $\delta \geq \Gamma$, $\omega=\sqrt{\delta^2-\Gamma^2}$,   $\cos\theta\equiv \Gamma/\sqrt{|\omega+\delta|^2+\Gamma^2}$ and $e^{i\varphi}\sin\theta \equiv -i (\omega+\delta)/\sqrt{|\omega+\delta|^2+\Gamma^2}$;
for $\delta < \Gamma$, the principal value is taken for $\omega =  i\kappa=i\sqrt{\Gamma^2-\delta^2}$.

\underline{\sectionn{Bell state generation.}}|We now show that steady Bell states can be generated
based on the monitored dynamics 
governed by $\bunderline{H}_{\text{eff}}$~\eqref{noneff}, focusing on the two-qubit dynamics in the subspace $\mathcal{H}$, which applies to zero and finite temperatures. Similarly to the unmonitored scenario, we can identify three distinct parameter regimes: parity-time ($\mathcal{PT}$) symmetry broken regime, $\delta<\Gamma$, the exceptional point, $\delta=\Gamma$, and $\mathcal{PT}$-exact regime, $\delta>\Gamma$~\cite{Heiss_2012,Galda2018prb,Galda2016prb}.

In the $\mathcal{PT}$-broken regime,  eigenvalues $E_{a}$ and $E_s$ are 
purely imaginary  with $\Im E_a>\Im E_s$.
Thus the probability in the eigenmode $\ket{a}$ ($\ket{s}$) grows (decays) in time,
and all probability eventually
flows
into the eigenmode $\ket{a}$.  
For an arbitrary initial state $\rho_0=\sum_{i,j=\{a,s\}}p_{ij}\ket{i}\!\!\bra{j}$, 
one can 
analytically solve for
$\rho(t)$ according to Eq.~\eqref{rhosol}:
  \(\rho(t)\! \!=\! \frac{ p_{aa}e^{2\kappa t} \! \ket{a}\!\!\bra{a} \!+\! p_{as}\ket{a}\!\!\bra{s} \!+\! p_{sa}\ket{s}\!\!\bra{a} \!+\! p_{ss}e^{-2\kappa t}\ket{s}\!\!\bra{s}   }{ p_{aa}e^{2\kappa t} +p_{as} \bra{s}\ket{a} +p_{sa}\bra{a}\ket{s} +p_{ss}e^{-2\kappa t}  } ,   \)
which ultimately evolves into the maximally entangled state $\ket{a}=\left(\ket{\uparrow\downarrow}+e^{i\tilde{\varphi}}\ket{\downarrow\uparrow}\right)/\sqrt{2}$ with $e^{i\tilde{\varphi}}=(-\kappa+i\delta)/\Gamma$, when $t\gg 1/2\kappa$. 
Thus, the two qubits eventually reach the maximal concurrence  in the $\mathcal{PT}$-broken regime,  irrespective of the initial state as long as $p_{aa}\neq 0$. 
As an illustrative example, we evaluate the time-dependent concurrence, with initial state $\ket{\uparrow\downarrow}$ and equal local fields $\delta=0$,  $  \mathcal{C}(t) =\tanh 2\Gamma t,$    as shown in Fig.~\ref{fig4}. 
In practice, 
the entanglement-growth rate $2\kappa$  needs to be larger than the postselection rate ($\sim$ the rate of leaking out of the subspace $\mathcal{H}$ of our interest) $2(a+\tilde{a}-\kappa)$ for the system to settle into the Bell state.
The optimal scenario is when $\delta = 0$,
$|A|=a$, and $|\tilde{A}|=\tilde{a}$, the same as that
in the overdamped quantum regime 
without postselection.

At the exceptional point,  $\bunderline{H}_{\text{eff}}$ is nondiagonalizable, since the eigenstates $\ket{a}$ and $\ket{s}$
coalesce into $\left( \ket{\uparrow\downarrow}+i\ket{\downarrow\uparrow} \right)/\sqrt{2}$. The two qubits will gradually evolve into this sole state where they are maximally entangled. For example, 
starting with a trivial state $\ket{\uparrow \downarrow}$, the  
concurrence 
$\mathcal{C}(t)=  2\Gamma t  \sqrt{1+\Gamma^2 t^2}  /( 1+2\Gamma^2 t^2 ), $
algebraically approaching 1.
See Fig.~\ref{fig4}.

In the $\mathcal{PT}$-exact regime, the eigenenergies $E_{s}$ and $E_a$ have nonzero real parts. The 
amplitudes of eigenmodes $\ket{s}$ and $\ket{a}$
keep oscillating without reaching a steady state,
hence no steady entanglement. The frequency of entanglement oscillation is $2\omega$, as shown in Fig.~\ref{fig4}. The maximal entanglement one can achieve is 
$   \mathcal{C}_{\text{max}}(\eta)=  \sqrt{2-  1/\eta^2}/\eta,  $ with $\eta=\delta/\Gamma$, which is less than 1.
Notably, the second derivative of $\mathcal{C}_{\text{max}}$ is discontinuous across the exceptional point ($\eta=1$), reflecting a  phase transition (see Fig.~\ref{fig4}). 

It is clear that we can achieve a Bell state by decreasing the local field asymmetry $\delta$ for a fixed dissipative coupling $\Gamma$ to reach the $\mathcal{P}\mathcal{T}$-broken regime. We remark that $\Gamma$ is also potentially tunable by engineering the magnetic medium spintronically.
This discussion, again, highlights 
the role of
the nonlocal dissipative couplings 
in realizing
an exceptional point in the dynamics, further triggering an entanglement transition in 
the long-time steady state.  In the large nonlocal dissipative coupling regime, we achieve steady Bell states.

\underline{\sectionn{Discussion.}}|We remark that the non-Hermitian Hamiltonian scheme is precise when the rate of measurements is infinite. As this rate approaches zero, we recover the full Lindblad dynamics.  It could be intriguing to explore, within our framework, possible  phase transitions or crossovers induced by finite-rate measurements. 

In our case, the possible forms of induced coherent interactions and quantum jump operators are determined by the axial symmetry of the  media. 
This may render a general guidance in studying the dynamics of hybrid quantum systems with other classes of symmetries, especially their long-time entanglement behavior.


The theoretical framework developed here provides a good starting point for further studies on the relationship between the entanglement dynamics and thermodynamic properties of the medium.
One may be able to manipulate   entanglement by engineering the  medium~\cite{cirac2004prl,cirac2011prl,Muschik_2012,McEndoo_2013}, enabled by recent progress in the field of spintronics~\cite{spincurrent22,Yaroslavreview,hectorrmp2020}. It is especially interesting to look into media with anisotropies, which have been shown to be good entanglement reservoirs~\cite{PhysRevB.101.014416, PhysRevB.100.174407, kamra2020apl}.  
By extending our equilibrium Green's function treatment to allow for a quasi-equilibrium spin chemical potential, we may  study the scenario with a spintronically pumped medium, where 
local relaxations and dissipative couplings are tunable.

For practical consideration in a NV-magnet hybrid setup, one challenge is to bring down the detuning $\delta$ of the magnetic fields at the NV sites to be much smaller than the dissipative parameters, which are typically on the scale of MHz or less. It would also be necessary to  put the induced effective Hamiltonian back into the picture, as the interplay between the coherent and dissipative evolution can be  nontrivial in steady-state entanglement generation. 
Though the Markovian nature of the intrinsic dynamics can be justified when the NV frequency $\Delta$ is sufficiently above the magnon band gap $\Omega$, such that the decay time of relevant correlations within the medium, $\sim (\Delta-\Omega)^{-1}$, is smaller than the timescale associated with the medium-induced qubit dynamics, disorder and defects may lead to low-energy excitations that contribute to non-Markovian evolution (which can also be interesting to look into).
 As a possible low-temperature implementation of the proposed post-selection scheme, we may post-select on the absence of any emitted magnons.

Lastly, NV centers have been demonstrated as good quantum probes of local fields and noise~\cite{Casola2018nature}. Here, we propose to extend this to nonlocal characteristics. For example, one may use it to detect quantum phase transitions and steady exotic phases that are characterized by nonlocal quantum correlations.

\begin{acknowledgements}
This work is supported by NSF under Grant No. DMR-2049979.

\end{acknowledgements}


%

\onecolumngrid
\clearpage
\setcounter{equation}{0}
\renewcommand{\theequation}{S\arabic{equation}}
\renewcommand{\thefigure}{S\arabic{figure}}
\appendix

{\centering
    \large{\textbf{{Supplemental Material for \\ ``Bell-state generation for spin qubits via dissipative coupling"}}}
\par}

\bigskip
\bigskip

\author{Ji Zou}
\email{jzeeb@ucla.edu}
\author{Shu Zhang}
\author{Yaroslav Tserkovnyak}
\affiliation{Department of Physics and Astronomy, University of California, Los Angeles, California 90095, USA}


\maketitle

In this Supplemental Material, we present (i) General formalism for open quantum systems, (ii) Single spin dynamics, (iii) Derivation of the master equation for the two-qubit scenario, (iv)  Symmetry-dictated possible forms of $H_{\text{eff}}$, (v)  Thermodynamic stability of the magnetic medium, (vi) Concurrence as a measure of entanglement, and (vii) Full entanglement dynamics.

\subsection*{(i) General formalism for open quantum systems}
A system exchanging information and energy with its environment can be modeled by
\(H=H_S+H_E+H_{\text{SE}},\)
where $H_S$ stands for the Hamiltonian of the system, $H_E$ is the Hamiltonian of the  environment, and $H_{\text{SE}}$ describes the  interactions.  
Denoting the total density matrix by $W(t)$, the reduce density matrix of the system is
$ \rho(t)=\Tr_E W(t)$, where $\Tr_E$ traces out the degrees of freedom of the environment.
We derive the master equation for $\rho(t)$ from
\(  \dv{\rho(t)}{t}=\Tr_E \dv{W(t)}{t}=-i\Tr_E[H,W(t)],  \label{sm4} \)
taking the Markov approximation 
\( W(t)\approx\rho(t)\otimes \rho_{\text{eq}}^E, \label{sm5} \)
where $\rho^E_{\text{eq}}=e^{-\beta H_E}/Z_E$ is the thermal equilibrium density matrix of the environment with $\beta=1/k_B T$. This approximation assumes the relaxation time of the environment is much faster than other time scales in the problem. 
In principle, the dynamics of the system will also affect the state of the environment, but the environment can relax to the thermal equilibrium very quickly due to the its large degrees of freedom, and, as a result, lose the memory of the changes.  In the magnetic systems, the decay time of  correlations within the medium is   $\hbar/(\Delta - \Omega)$ with $\Delta$ being the typical energy scale of $H_S$ and $\Omega$ being the magnon band gap, which physically is the time it takes for the magnon of frequency $\Delta/\hbar$ to propagate over  the distance set by its own wavelength. Thus the Markov approximation is justified when the energy scale of the system $\Delta$ is large enough compared with the magnon gap, such that the medium relaxation time is smaller than the time scale associated with the medium-induced dynamics of the system. 
Assuming system-bath coupling is small, we take the Born approximation
 (expanding Eq.~(\ref{sm4}) to the second order) and also make the Markov approximation (so we get a master equation that is local in time), which yields the master equation for $\rho(t)$ in the Schrödinger picture~\cite{Heinz, Lidar2019}:
\(
\dot{\rho}(t)=-i[H_S, \rho(t)]-\int^\infty_0d\tau \,\, \Tr_E\Big\{  \big[  H_{\text{SE}},[H_{\text{SE}}(-\tau),\rho(t)\otimes \rho^{\text{eq}}_E]   \big]    \Big\},  \label{sm8}  \)
where $H_{\text{SE}}(-\tau)$ is in the interaction picture and we have set $\hbar=1$ for simplicity. This  expression is valid for a general interaction
$ H_{\text{SE}}=\sum_\alpha S_\alpha\otimes B_\alpha  $,
with $\langle B_\alpha\rangle_{\text{eq}}=0$. Here, $S_\alpha$ and $B_\alpha$ are operators acting on Hilbert spaces of the system and the environment, respectively. For $\langle B_\alpha\rangle_{\text{eq}}\neq 0$, we can always redefine $ B \rightarrow B- \langle B\rangle_{\text{eq}}$. 
The second term in Eq.~(\ref{sm8}) can be written as 
\(      \Tr_E\Big\{  \big[  H_{\text{SE}},[H_{\text{SE}}(-\tau),\rho(t)\otimes \rho^{\text{eq}}_B]   \big]    \Big\}= \langle H_{\text{SE}}H_{\text{SE}}(-\tau)\rangle_{\text{eq}} \rho - \langle H_{\text{SE}}(-\tau) \rho H_{\text{SE}} \rangle_{\text{eq}}+\text{H.c.}.  \label{key} \)
It is straightforward to work out the master equation for $\rho(t)$ once the interaction is specified. 

\subsection*{(ii) Single spin dynamics}
In this section, we derive the well-known relaxation time $T_1$ and decoherence time $T_2$, which characterize the local relaxation of a single qubit, by considering the dynamics of a spin-$1/2$ system $H_S=-\Delta\sigma^z/2$  coupled to a bath $H_E$. 
A general form of the interaction can be in the first order of the Pauli matrices $\sigma^z$ and $\sigma^\pm = (\sigma^x \pm i \sigma^y)/2$ for the spin-$1/2$,
$H_{\text{SE}}=\sigma^z\otimes X+ \sigma^+\otimes Y+\sigma^-\otimes Y^\dagger$, with operators $X$ and $Y$ acting on the bath. We assume $X = X^\dagger$, and $\langle X\rangle =\langle Y\rangle =0$, where the average denotes the thermal average taken within the bath.  In the interaction picture,
\( H_{\text{SE}}(-\tau) = \sigma^z\otimes X(-\tau)+\sigma^+\otimes Y(-\tau)e^{i\Delta \tau} +\sigma^-\otimes Y^\dagger(-\tau) e^{-i\Delta\tau}, \)
and Eq.~(\ref{key}) then becomes
\(  \langle XX(-\tau)\rangle (\sigma^z\sigma^z\rho -\sigma^z\rho \sigma^z) +   \langle YY^\dagger(-\tau)\rangle e^{-i\Delta \tau} (\sigma^+\sigma^-\rho - \sigma^- \rho \sigma^+)  + \langle Y^\dagger Y(-\tau)\rangle e^{i\Delta \tau} (\sigma^-\sigma^+\rho - \sigma^+ \rho \sigma^-)  +\text{H.c.}  \label{onespin}  \)
where we have assumed certain symmetry properties of the bath such that only $\langle XX\rangle, \langle YY^\dagger \rangle$, and $\langle Y^\dagger Y\rangle$ are nonzero. For example, taking $X = S^z$ and $Y = S^-$ to be spin operators in the bath, the $U(1)$ symmetry in the spin space of the bath is assumed.
We therefore obtain the master equation:
\( \dot{\rho}=-i\big[H_S + H_{\text{Lamb}},\rho\big]-\mathcal{L}[\rho], \label{12}  \)
where  the Lamb shift is 
\( H_{\text{Lamb}}= \frac{ {\rm Im} \int^\infty_0d\tau \langle \{Y(\tau) ,Y^\dagger \} \rangle e^{-i\Delta \tau}}{2} \sigma^z, \label{lam}   \)
and the Lindbladian superoperator is given by: 
 \eq{\al{ \mathcal{L}[\rho]&=  \frac{1}{2}\int^\infty_{-\infty}d\tau \langle X(\tau)X\rangle (2\rho-2\sigma^z\rho \sigma^z)  \\  
 &+  \frac{1}{2}\int^\infty_{-\infty} d\tau \langle Y(\tau) Y^\dagger \rangle e^{-i\Delta \tau} \; \Big(  \sigma^+\sigma^-\rho +\rho \sigma^+\sigma^- -2 \sigma^-\rho \sigma^+ \Big)  \\
  &+ \frac{1}{2}\int^\infty_{-\infty} d\tau \langle Y^\dagger Y(\tau) \rangle e^{-i\Delta \tau} \; \Big(  \sigma^-\sigma^+\rho +\rho \sigma^-\sigma^+ -2 \sigma^+\rho \sigma^- \Big). }}
For notational convenience, let us introduce the following parameters to denote the coefficients in the Lindbladian:
  \eq{\al{ 2D&\equiv \int^\infty_{-\infty}d\tau \langle X(\tau)X\rangle=iG^>_{XX}(\omega=0)=iG^<_{XX}(0)= \frac{iG^>_{XX}(0)+iG^<_{XX}(0)}{2}= \frac{S_X(0)}{2}, \\
     2B^<&\equiv \int^\infty_{-\infty} d\tau \langle Y(\tau) Y^\dagger \rangle e^{-i\Delta \tau} =iG^<_{Y^\dagger Y}(\Delta), \\
      2B^>&\equiv \int^\infty_{-\infty} d\tau \langle Y^\dagger Y(\tau) \rangle e^{-i\Delta \tau} = iG^>_{Y^\dagger Y} (\Delta),  \\
      2\Bar{B}&\equiv B^>+B^<= \frac{iG^>_{Y^\dagger Y}(\Delta)+iG^<_{Y^\dagger Y}(\Delta)}{2}=\frac{S_{Y^\dagger}(\Delta)}{2}, 
      }\label{eq:single-qubit-supp}}
where \textit{greater} and \textit{lesser} Green's functions follow the conventional definition:
  \( G^>_{X,Y}(t)\equiv -i \langle X(t)Y\rangle, \;\;\; G^<_{X,Y}(t)=-i\langle YX(t)\rangle.  \)
  We have also introduce the equilibrium symmetrized fluctuations
\( iG^>_{X,Y}(\omega)+iG^<_{X,Y}(\omega)=\int dt \langle \{X(t),Y\}\rangle e^{i\omega t},  \)
specifically, we have the power spectrum 
\( S_{X}(\omega)=\int dt \langle \{X(t),X^\dagger\}\rangle e^{i\omega t}. \)
Here, all parameters $D, B^>, B^<, \Bar{B}$ are real-valued by their definitions. Furthermore, being the physical decay rates, they must be non-negative, which is ensured by the thermodynamic stability of the bath, as will be detailed later. 

According to the master equation \eqref{12}, we obtain the equation of motion for $s^z = \langle \sigma^z \rangle/2 = (\rho_{11} - \rho_{22})/2$:
\( \dv{}{t} s^z = -4\Bar{B}\big(  s^z -\bar{s}^z  \big), \label{sz} \)
from which we extract the relaxation time $T_1$: 
\( T_1^{-1}=4\bar{B}=S_{Y^\dagger}(\Delta),  \)
where
$\bar{s}^z\equiv (B^>-B^<)/4\bar{B}$ is the equilibrium value of the spin-$z$ component.
This value can be made sense of by writing down the equation for the diagonal elements of the density matrix: 
\(\label{eq:diagnoal-drho-dt} 
\dot{\rho}_{11}=-2B^<\rho_{11}+2B^>\rho_{22},  
\quad \text{and} \quad
 \dot{\rho}_{22}=2B^<\rho_{11}-2B^>\rho_{22}, \)
which implies that the transition rate of flipping the spin from up to down is $2B^<$ and that of the reverse process is $2B^>$. In equilibrium,
\( \frac{\rho_{11}}{\rho_{22}}=\frac{B^>}{B^<}, \)
from which we conclude that the probability of measuring $\sigma^z$ to be $1$ is $B^>/2\bar{B}$ and $-1$ is $B^</2\bar{B}$, which satisfies $\rho_{11}+\rho_{22}=1$. 
We next look into the time evolution of the off-diagonal elements: 
\(\label{eq:off-diagnoal-drho-dt} 
\dv{}{t}\rho_{12}=i\Delta \rho_{12}-(4D+2\bar{B})\rho_{12},\quad
\text{and} \quad \dv{}{t}\rho_{21}=-i\Delta \rho_{21}-(4D+2\bar{B})\rho_{21}. \)
Since $ s^x = \langle \sigma^x \rangle =(\rho_{12}+\rho_{21})/2$ and $s^y = \langle \sigma^y \rangle =i(\rho_{12}-\rho_{21})/2$,
\(  \dv{}{t} s^x =\Delta  s^y-(4D+2\bar{B}) s^x\quad \text{and} \quad \dv{}{t} s^y =-\Delta s^x-(4D+2\bar{B}) s^y. \label{sxy} \)
We thus identify the decoherence rate: 
\( T_2^{-1}=4D+2\bar{B}=S_X(0)+\frac{S_{Y^\dagger}(\Delta)}{2}.  \)
Note that $S_X(0)$ contributes to the dephasing effect only.
This is rooted in the fact that $\sigma^z\otimes X$ commutes with the single NV center Hamiltonian $\Delta \sigma^z/2$. As a result, this type of interaction does not lead to energy flow between the system and the bath, but information flow only, i.e. a relative phase damping  between NV center levels. 

Combining Eq.~(\ref{eq:diagnoal-drho-dt}) and Eq.~(\ref{eq:off-diagnoal-drho-dt}), the equations of motion for the density matrix can be put into the following form:
\( \dv{}{t}\hat{\rho}=\mqty(  -2B^<\rho_{11}+2B^>\rho_{22} & i\Delta \rho_{12} -T_2^{-1}\rho_{12} \\  -i\Delta \rho_{21}-T_2^{-1} \rho_{21} & 2B^<\rho_{11}-2B^>\rho_{22}  ).  \)

We remark that Eq.~(\ref{sz}) and Eq.~(\ref{sxy}) are nothing but the Bloch equations: 
\eq{\al{\dv{ s^x}{t} &=\big(  \Vec{s}  \times \Vec{B}\big)_x - \frac{ s^x}{T_2},\\
\dv{ s^y}{t} &=\big(  \Vec{s}  \times \Vec{B}\big)_y - \frac{ s^y}{T_2},\\
\dv{ s^z}{t} &=\big(  \Vec{s}  \times \Vec{B}\big)_z - \frac{ s^z-\bar{s}^z}{T_1},
}}
with the magnetic field  $\Vec{B}=\Delta \hat{z}$.

\subsection*{(iii) Derivation of the master equation for the two-qubit scenario}
In this section, we derive the master equation [Eq.~(\textcolor{red}{2}) in the main text] for the two-qubit system interacting with a magnetic medium. The system Hamiltonian is $H_S=-( \Delta_1\sigma_1^z + \Delta_2\sigma_2^z  )/2$ and the  interaction Hamiltonian is
$H_{\text{SE}}=\lambda\sum_{\alpha = 1,2} \big( \sigma_\alpha^+ S_\alpha^-  +\sigma_\alpha^- S_\alpha^+ +\sigma_\alpha^z S^z_\alpha \big)$,
where $\sigma^\pm=(\sigma^x \pm i\sigma^y)/2$, $S^{\pm}=S^x\pm iS^y$, and $\vec{S}_\alpha$ is  the local spin density operator within the magnetic medium. 
To apply the formalism~(\ref{sm8}) directly, we shift the total Hamiltonian by $-\lambda \langle S^z \rangle (\sigma^z_1+\sigma_2^z)$, which is equivalent to making $S^z_\alpha\rightarrow \Tilde{S}^z_\alpha = S^z_\alpha -\langle S^z\rangle$, such that $\langle \Tilde{S}^z_\alpha \rangle=0$. 
Here and in what follows, we refer to $\Tilde{S}^z$ as $S^z$ for notational convenience. Without loss of generality, we assume $\Delta_1\geq \Delta_2\geq 0$.

Substituting  
\( H_{\text{SE}}(-\tau)=\lambda\sum_\alpha \Big[  \sigma_\alpha^+S^-_\alpha(-\tau) e^{i\Delta_\alpha \tau} +\sigma_\alpha^- S^+_\alpha(-\tau) e^{-i\Delta_\alpha \tau} +\sigma_\alpha^z S^z(-\tau)   \Big],  \)
into Eq.~\eqref{key} yields an explicit expression 
\( \langle S^-_\alpha S^+_\beta(-\tau)\rangle e^{-i\Delta_\beta \tau} ( \sigma^+_\alpha \sigma^-_\beta \rho -\sigma^-_\beta \rho \sigma^+_\alpha  )+  \langle S^+_\alpha S^-_\beta(-\tau)\rangle e^{i\Delta_\beta \tau} ( \sigma^-_\alpha \sigma^+_\beta \rho -\sigma^+_\beta \rho \sigma^-_\alpha  ) + \langle S^z_\alpha S^z_\beta(-\tau) (\sigma^z_\alpha \sigma^z_\beta \rho -\sigma_\beta^z \rho\sigma_\alpha^z)  +  \text{H.c.} , \)
where Einstein summation is implied over $\alpha,\beta = 1,2$.   The effect of the above terms on the evolution of the density matrix can be grouped under two operators, an effective Hamiltonian and a Lindbladian, in the master equation~(\ref{sm8}):
\(  \dot{\rho}=-i[H_S+H_{\text{eff}},\rho]-  \mathcal{L}[\rho].  \)
The medium-induced effective qubit-qubit interaction is
\[ H_{\text{eff}}= \mathcal{J}_1^{\alpha\beta}\sigma_\alpha^+\sigma_\beta^- + \mathcal{J}_2^{\alpha\beta} \sigma^-_\alpha \sigma_\beta^++ \mathcal{J}_z^{\alpha\beta} \sigma^z_\alpha \sigma^z_\beta, 
\label{eq:coherent-coupling-supp}\]
and the Lindbladian superoperator is given by
\eq{\mathcal{L}[\rho]=\tilde{A}_{\alpha\beta} (\sigma_\alpha^+\sigma_\beta^-\rho +\rho \sigma_\alpha^+\sigma_\beta^- -2 \sigma_\beta^-\rho \sigma^+_\alpha)  + A_{\alpha\beta} (\sigma_\alpha^-\sigma_\beta^+\rho +\rho \sigma_\alpha^-\sigma_\beta^+ -2 \sigma_\beta^+\rho \sigma^-_\alpha)  +D_{\alpha\beta} ( \sigma^z_\alpha \sigma_\beta^z \rho +\rho \sigma_\alpha^z \sigma_\beta^z -2 \sigma_\beta^z \rho \sigma_\alpha^z   ).  \label{smlind} }

\textbf{\textit{Coherent couplings}.}|The coherent couplings in Eq.~(\ref{eq:coherent-coupling-supp}) can all be related to the real part of retarded Green's functions of the magnetic medium. Firstly,
\( \mathcal{J}_z^{\alpha\beta}=- \frac{i\lambda^2}{2} \int^\infty_0 \big[ \langle S^z_\alpha S^z_\beta (-\tau) \rangle - \langle S^z_\alpha (-\tau) S^z_\beta \rangle  \big]d\tau =-\frac{i\lambda^2}{2}\int^\infty_0 \langle [S^z_\alpha(\tau),S^z_\beta]\rangle d\tau =\lambda^2\frac{G^R_{S^z_\alpha S^z_\beta}(0)}{2}  \)
is real valued by definition.
Here we have adopted the standard definition for the retarded Green's function  $G^R_{AB}(\omega)=-i\int^\infty_0 \langle[A(\tau),B] \rangle e^{i\omega \tau}d\tau$.
Secondly,
\eq{\al{ \mathcal{J}_1^{\alpha\beta}&=\frac{-i}{2}\int^\infty_0 \big[ \langle S^-_\alpha S^+_\beta (-\tau) \rangle e^{-i\Delta_\beta \tau} - \langle S^-_\alpha(-\tau) S^+_\beta \rangle e^{i\Delta_\alpha \tau} \big] d\tau,  \\
  \mathcal{J}_2^{\alpha\beta} &=\frac{-i}{2} \int^\infty_0 \big[  \langle S^+_\alpha S^-_\beta(-\tau) \rangle e^{i\Delta_\beta \tau} -\langle S^+_\alpha(-\tau) S^-_\beta\rangle e^{-i\Delta_\alpha\tau} \big] d\tau.    }}
Therefore,
\eq{\al{ \mathcal{J}_1^{\alpha\beta}\sigma_\alpha^+\sigma_\beta^- + \mathcal{J}_2^{\alpha\beta} \sigma^-_\alpha \sigma_\beta^+ =& \sum_{\alpha\neq \beta} ( \mathcal{J}_1^{\alpha\beta} +\mathcal{J}_2^{\beta\alpha}  )\sigma_\alpha^+\sigma_\beta^-+ \sum_\alpha \big( \mathcal{J}_1^{\alpha\alpha}\sigma_\alpha^+\sigma_\alpha^-+\mathcal{J}_2^{\alpha\alpha}\sigma_\alpha^-\sigma_\alpha^+  \big) \\
=& \sum_{\alpha\neq \beta} ( \mathcal{J}_1^{\alpha\beta} +\mathcal{J}_2^{\beta\alpha}  )\sigma_\alpha^+\sigma_\beta^-+ \sum_\alpha \frac{\mathcal{J}_1^{\alpha\alpha}-\mathcal{J}_2^{\alpha\alpha}}{2} [\sigma_\alpha^+,\sigma_\alpha^-] +\sum_\alpha \frac{\mathcal{J}_1^{\alpha\alpha}+\mathcal{J}_2^{\alpha\alpha}}{2} (\sigma_\alpha^+\sigma_\alpha^-+\sigma_\alpha^-\sigma_\alpha^+) \\
=& \lambda^2 \frac{G^R_{S_2^+S_1^-}(\Delta_1)+G^A_{S^+_2S^-_1}(\Delta_2)}{2} \sigma_1^+\sigma_2^-+\text{H.c.} \\
&  +\sum_\alpha \Big[\lambda^2{\rm Im}\int^\infty_0\langle \{ S^-_\alpha(\tau), S_\alpha^+ \}\rangle e^{-i\Delta_\alpha \tau} d\tau  \Big] \frac{\sigma_\alpha^z}{2} \\
& +\lambda^2  \sum_\alpha \frac{  G^R_{S^+_\alpha S^-_\alpha}(\Delta_\alpha) +G^A_{S^+_\alpha S^-_\alpha}(\Delta_\alpha)     }{4}     \\
\approx&\big[\lambda^2 \mathfrak{Re} G^R_{S_2^+ S_1^-}(\Delta) \sigma_1^+\sigma_2^- +\text{H.c.} \big] + H_{\text{Lamb}} +\text{const.} }}
is Hermitian.
In the last step, we have approximated $\Delta_1 \approx \Delta_2 \approx \Delta\equiv (\Delta_1+\Delta_2)/2$. This is sufficient for the discussion of medium-induced dynamics (both coherent and dissipative), since we are primarily interested in the scenario with $\Delta_1 \approx \Delta_2$, on the scale set by induced interqubit coupling $\propto\lambda^2$. The expansion with respect to $\Delta_1 - \Delta_2$ would thus induce higher-order corrections. Here, we have denoted $\mathfrak{Re}\, G^R_{A,B}(t,t^\prime)\equiv\big[ G^R_{A,B}(t,t^\prime)+G^A_{A,B}(t,t^\prime) \big]/2$, noting this ``real" part becomes the ordinary real part $\mathfrak{Re} \rightarrow \Re$ when $A=B^\dagger$. The Lamb shift $H_{\text{Lamb}}\propto \sigma^z$, which we have already encountered  at the level of single-spin dynamics Eq.~\eqref{lam}, can be absorbed into the bare system Hamiltonian $H_S$, and we drop the real constant term in later discussions. We therefore obtain the total induced effective interaction Hamiltonian~(\ref{eq:coherent-coupling-supp}): 
\(  H_{\text{eff}}/\lambda^2 = \big[\mathfrak{Re} G^R_{S_2^+S_1^-} (\Delta) \, \sigma_1^+\sigma_2^-+\text{H.c.} \big]+  \frac{G^R_{S_1^zS_2^z}(0)  +  G^R_{S_2^zS_1^z}(0)  }{2}    \, \sigma_1^z\sigma_2^z,  \label{smheff}  \)
which is a close analogy to the RKKY interactions induced by itinerant electrons between magnetic moments in a metal. 

\textbf{\textit{Lindbladian}.}|We now derive Eq.~(\textcolor{red}{4}-\textcolor{red}{6}) in the  main text to identify the two types of dissipation|local relaxation and dissipative couplings|expressed in terms of greater or less Green's functions of the magnet. They are related to the imaginary part of retarded Green's functions via \textit{the fluctuation-dissipation theorem}. The coefficients in the Lindbladian operator~\eqref{smlind} are given by
\eq{\al{ \tilde{A}_{\alpha\beta}&=\frac{\lambda^2}{2}\int^\infty_0 \big[  \langle S^-_\alpha S^+_\beta(-\tau)\rangle e^{-i\Delta_\beta \tau} + \langle S^-_\alpha (-\tau) S^+_\beta \rangle e^{i\Delta_\alpha \tau} \big] d\tau \approx \frac{\lambda^2}{2} \int^\infty_{-\infty} \langle S^-_\alpha S^+_\beta(\tau) \rangle e^{i\Delta \tau} d\tau \equiv \frac{i\lambda^2 G^<_{S^+_\beta S_\alpha^-}(\Delta)}{2} , \\
 A_{\alpha\beta} &= \frac{\lambda^2}{2} \int^\infty_0 \big[    \langle S^+_\alpha S^-_\beta (-\tau) \rangle e^{i\Delta_\beta\tau} + \langle S^+_\alpha(-\tau)S^-_\beta \rangle e^{-i\Delta_\alpha\tau} \big] d\tau \approx \frac{\lambda^2}{2} \int^\infty_{-\infty} \langle S^+_\alpha(\tau) S^-_\beta \rangle  e^{i\Delta \tau}  d\tau   \equiv \frac{i \lambda^2 G^>_{S^+_\alpha S_\beta^-}(\Delta)}{2}   , \\
 D_{\alpha\beta} &=\frac{\lambda^2}{2} \int^\infty_0 \big[ \langle S^z_\alpha S^z_\beta(-\tau) \rangle + \langle S^z_\alpha(-\tau) S^z_\beta \rangle \big] d\tau = \frac{\lambda^2}{2} \int^\infty_{-\infty} \langle S^z_\alpha (\tau) S^z_\beta \rangle d\tau\equiv \frac{i\lambda^2  G^>_{S^z_\alpha S^z_\beta}(0)}{2} = \frac{i\lambda^2 G^<_{S^z_\beta S^z_\alpha}(0)}{2} , }}
 where the approximation $\Delta_1 \approx \Delta_2 \approx \Delta$ is again taken in the first two equations.  It can be directly observed that $A_{\alpha\alpha},\tilde{A}_{\alpha\alpha}, D_{\alpha\alpha}$ are real valued, and $A_{12}^*=A_{21},\tilde{A}_{12}^*=\tilde{A}_{21},D_{12}^*=D_{21}$. We introduce the following parameters to clean up the notation: $a\equiv A_{11}=A_{22}$, $\tilde{a}\equiv \tilde{A}_{11}=\tilde{A}_{22}$,  $d\equiv D_{11}=D_{22}$, assuming the medium is homogeneous, and $A\equiv A_{12}$,  $\tilde{A}\equiv \tilde{ A}_{21}$,  and $\mathfrak{D}\equiv D_{12}$. We then can write the Lindbladian \eqref{smlind} into the form of
 \(  \mathcal{L}[\rho]= \sum_{nm}h_{nm}  \big(  \mathcal{O}^\dagger_m \mathcal{O}_n \rho + \rho \mathcal{O}^\dagger_m \mathcal{O}_n -2 \mathcal{O}_n \rho \mathcal{O}^\dagger_m   \big), \label{smh}   \) 
with  $\mathcal{O}\!\! =\!\! ( \sigma_1^- , \sigma_2^- , \sigma_1^+ , \sigma_2^+ , \sigma_1^z , \sigma_2^z )$ and 
\(  h= \mqty(\tilde{a} & \tilde{A} \\ \tilde{A}^* & \tilde{a}) \oplus  \mqty(a & A^* \\ A & a) \oplus \mqty(d & \mathfrak{D}\\  \mathfrak{D}^* &d),  \) 
where $\oplus$ denotes the direct sum of matrices.
Originated from those terms in $H_{\text{SE}}$ that commute with $H_{\text{S}}$, namely $\lambda\sum_{i=1,2}\sigma_i^zS^z_i$, the terms in Eq.~(\ref{smh}) with parameters $d$ and $\mathfrak{D}$ are pure dephasing. They only cause information but not energy exchange between the system and the medium, and can be suppressed by dynamic decoupling in practice. 

In the following discussion, we focus on the four remaining dissipative parameters: $a$, $\tilde{a}$, $A$, and $\tilde{A}$. 
For one qubit, for example, by setting $\alpha, \beta=1$, we reproduce the results of the single qubit scenario, where the local relaxation of individual qubits is governed by the parameters $a$ and $\tilde{a}$. They correspond to $B^<$ and $B^>$ in Eq.~(\ref{eq:single-qubit-supp}) associated with local decay and the reverse process.
The parameters $A$ and $\tilde{A}$ apply to a system with multiple qubits and are related to cooperative decay and the reverse process, which we refer to as dissipative couplings. These parameters are not independent of each other: $\tilde{a}=e^{-\beta \Delta} a$ and $\tilde{A}=e^{-\beta \Delta}A$, in thermal equilibrium, where $\beta=1/k_BT$. In particular at $T=0$, the decay processes $\tilde{a}=\tilde{A}=0$. We remark that one can generalize these relations even when the magnetic medium is pumped out of equilibrium, for example, quasi-equilibrium where magnons have a finite chemical potential.

The coefficient matrix $h$ is diagonalizable in general and Eq.~\eqref{smh} can take the form of 
 \( \underline{\mathcal{L}}[\rho] =    \sum_{i=1}^4 \mathcal{D}_{J_i}[\rho] ,   \)
with the dissipator $\mathcal{D}_J[\rho]\equiv J^\dagger J \rho + \rho J^\dagger J -2J \rho J^\dagger $. There are four quantum-jump operators (as we have neglected the dephasing effects):
\(  J_1= \sqrt{\frac{\tilde{a}+|\tilde{A}|}{2}}    \left( \sigma_1^-+\sigma_2^-  \right),  J_2=  \sqrt{\frac{\tilde{a}-|\tilde{A}|}{2}}     \left( \sigma_1^--\sigma_2^-  \right),  
    J_3= \sqrt{\frac{a+|A|}{2}}  \left( \sigma_1^++\sigma_2^+  \right),  J_4=    \sqrt{\frac{a-|A|}{2}}   \left( \sigma_1^+-\sigma_2^+  \right).  \label{smjump}   \)

\subsection*{(iv)  Symmetry-dictated possible forms of $H_{\text{eff}}$}
For a two-qubit system with the axial symmetry around $z$ axis in spin space, the interaction between the two qubits in general takes the form of
\( H_{\text{eff}}=\mathcal{J}_z\sigma^z_1\sigma^z_2+\mathcal{J}_\perp (\sigma_1^x\sigma_2^x  + \sigma_1^y\sigma_2^y)+D \hat{z}\cdot \vec{\sigma}_1\times \vec{\sigma}_2,  
\label{eq:two-qubit-interaction-supp}\)
which is an XXZ model allowing a Dzyaloshinskii-Moriya (DM) interaction. 
Further, adding a mirror symmetry with respect to the $xz$-plane (containing both qubit sites) dictates a vanishing $D$. In the fully isotropic limit, the interaction becomes Heisenberg $H_{\text{eff}}\propto \vec{\sigma}_1\cdot \vec{\sigma}_2$.

Denoting $\mathfrak{Re}G^R_{S_2^+S_1^-} (\Delta)=g_1+ig_2$,
\( \mathfrak{Re} G^R_{S_2^+S_1^-} (\Delta) \, \sigma_1^+\sigma_2^-+\text{H.c.} =     \frac{g_1}{2} (\sigma_1^x\sigma_2^x  + \sigma_1^y\sigma_2^y)  + \frac{g_2}{2} \hat{z} \cdot \vec{\sigma}_1\times \vec{\sigma}_2.   \)
We precisely obtain the effective two-qubit Hamiltonian~\eqref{smheff} in the general form~(\ref{eq:two-qubit-interaction-supp}), where we identify $\mathcal{J}_z= [G^R_{S_1^zS_2^z}(0)  +  G^R_{S_2^zS_1^z}(0) ]  /2$, $\mathcal{J}_\perp=g_1/2$, and $D=g_2/2$, setting the irrelevant overall factor $\lambda = 1$. 
For the explicit expressions:
\eq{\al{g_1&= \frac{  G^R_{S^+_2 S^-_1} (\Delta) + G^R_{S^+_1 S^-_2} (\Delta) +  G^A_{S^+_2 S^-_1} (\Delta) + G^A_{S^+_1 S^-_2} (\Delta)  }{4}, \\
  g_2&= \frac{ G^R_{S^+_2 S^-_1} (\Delta) - G^R_{S^+_1 S^-_2} (\Delta) +  G^A_{S^+_2 S^-_1} (\Delta) - G^A_{S^+_1 S^-_2} (\Delta)    }{4i}. }}
Both $g_1$ and $g_2$ are real, $g_1$ is nonzero in general, and $g_2$ is finite only when the $xz$-reflection symmetry as well as the $\pi$ $x$- or $z$-rotation symmetries are broken in the system. The DM interaction, similar to other coherent couplings, can build up finite entanglement between the two qubits. For example, taking $H_{\text{DM}}=D\hat{z}\cdot \vec{\sigma}_1\times \vec{\sigma}_2$ and initial state $\ket{\uparrow\downarrow}$, we obtain for the state of the two qubits:   $\ket{\psi(t)}=\cos(2Dt) \ket{\uparrow\downarrow} -\sin(2Dt) \ket{\downarrow\uparrow}$, which has concurrence $\mathcal{C}_{\text{DM}}(t)=|\sin(4Dt)|$. Therefore, we can conclude that the entanglement would oscillate between 0 and 1 with frequency $8D$.  We typically require the timescale $1/D$ to be shorter than the timescale set by the dissipation such that we can make use of the entanglement before it decays to zero.

In the SU(2)-symmetric limit and setting $\Delta \rightarrow 0$, 
the effective Hamiltonian~\eqref{smheff} simplifies into the Heisenberg form, as expected:
\( H_{\text{eff}} = \frac{ G^R_{S_2^x S_1^x}(0) +G^R_{S_1^x S_2^x}(0)  }{2} \,  ( \sigma_1^x\sigma_2^x  + \sigma_1^y\sigma_2^y  ) + \frac{ G^R_{S_2^zS_1^z}(0) + G^R_{S_1^zS_2^z}(0) }{2}    \, \sigma_1^z\sigma_2^z =\frac{   G^R_{S_2^x S_1^x}(0) +G^R_{S_1^x S_2^x}(0)   }{2}  \;  \vec{\sigma}_1\cdot \vec{\sigma}_2. \)
If, furthermore, the qubit sites can be exchanged under a spatial symmetry, then $H_{\text{eff}}=G^R_{S_1^xS_2^x}(0) \, \vec{\sigma}_1\cdot \vec{\sigma}_2$.

\subsection*{(v)  Thermodynamic stability of the magnetic medium}
The positive semidefinite evolution governed by Eq.~(\textcolor{red}{2}) in the main text requires the constraints:
\(a\geq |A|, \;\;\;\;\; \tilde{a}\geq |\tilde{A}|.  \)
In this section, we show that they are naturally guaranteed by the thermodynamic stability of the magnetic medium. In the following, we show $\tilde{a}\geq |\tilde{A}|$, namely $ iG^<_{S^+S^-}(\omega) \geq \big| iG^<_{S^+_2S^-_1}(\omega)  \big|$, and $a\geq |A|$ can be proved in the same spirit. Let us consider the response in the magnetic medium to the following perturbation:
\( H^\prime (t)=\sum_{\alpha=1,2} \big( \lambda_\alpha e^{-i\omega t} S^-_\alpha +\text{H.c.} \big)  =\lambda_1 e^{-i\omega t} S^-_1 +\lambda_1^* e^{i\omega t} S_1^+ +\lambda_2 e^{-i\omega t} S_2^- +\lambda_2^* e^{i\omega t} S_2^+.     \)
The corresponding linear-response dissipation power can be calculated via
\eq{\al{ P\equiv&\dv{}{t}\langle H(t)\rangle =\langle \partial_t H^\prime(t)\rangle = -i\omega e^{-i\omega t} \sum_{\alpha=1,2} \lambda_\alpha \langle S^-_\alpha \rangle +i\omega e^{i\omega t} \sum_{\alpha=1,2} \lambda_\alpha^*\langle S^+_\alpha \rangle \\
 =& -i\omega \sum_{\alpha,\beta} \lambda_\alpha \frac{\lambda^*_\beta}{\hbar} G^A_{S^+_\beta S^-_\alpha}(\omega) + i \omega \sum_{\alpha, \beta} \lambda_\alpha^* \frac{\lambda_\beta}{\hbar} G^R_{S^+_\alpha S^-_\beta} (\omega) \\
  =& i\omega \frac{|\lambda_1|^2}{\hbar} \big[ G^R_{S^+_1S^-_1}(\omega) - G^A_{S^+_1S^-_1}(\omega)  \big] + i\omega \frac{|\lambda_2|^2}{\hbar} \big[ G^R_{S^+_2S^-_2}(\omega) - G^A_{S^+_2S^-_2}(\omega)  \big] \\
  & +i\omega \frac{\lambda_1\lambda_2^*}{\hbar}\big[ G^R_{S^+_2S^-_1}(\omega) - G^A_{S^+_2S^-_1}(\omega)  \big] +i\omega \frac{\lambda_1^*\lambda_2}{\hbar} \big[ G^R_{S^+_1S^-_2}(\omega) - G^A_{S^+_1S^-_2}(\omega)  \big]  \\
  =& i\omega \frac{|\lambda_1|^2}{\hbar} 2i\Im G^R_{S^+_1S^-_1}(\omega) + i\omega \frac{|\lambda_2|^2}{\hbar} 2i\Im G^R_{S^+_2S^-_2}(\omega) + i\omega \frac{\lambda_1\lambda_2^*}{\hbar} 2i \,  \mathfrak{Im}\, G^R_{S^+_2S^-_1}(\omega) + i\omega \frac{\lambda_1^*\lambda_2}{\hbar} 2i \,  \mathfrak{Im}\, G^R_{S^+_1S^-_2}(\omega) \\
  =& \frac{\omega |\lambda_1|^2}{\hbar} A_{S^+_1S^-_1}(\omega) + \frac{\omega |\lambda_2|^2}{\hbar} A_{S^+_2S^-_2}(\omega) +  \frac{\omega \lambda_1 \lambda_2^* }{\hbar} A_{S^+_2S^-_1}(\omega)  +  \frac{\omega \lambda_1^* \lambda_2 }{\hbar} A_{S^+_1S^-_2}(\omega),      }}
where
we have used the Kubo formula in
\(  \langle S^+_\alpha\rangle
= \sum_{\beta} \frac{\lambda_\beta}{\hbar} e^{-i\omega t} G^{R}_{S^+_\alpha S^-_\beta}(\omega),\;\;\text{and}\;\;  
\langle S^-_\alpha\rangle =\langle S^+_\alpha\rangle^*=\sum_{\beta={1,2}}\frac{\lambda_\beta^*}{\hbar} e^{i\omega t} G^A_{S^+_\beta S^-_\alpha}(\omega), \)
Here, $\mathfrak{Im}G^R_{AB}(\omega)=\big[ G^R_{AB}(\omega) -G^A_{AB}(\omega) \big]/2i$ is the ``imaginary" part of the retarded Green's function, which is reduced to the ordinary imaginary part for $A=B^\dagger$, and $A_{A,B}(\omega)\equiv -2\mathfrak{Im}G^R_{A,B}(\omega)=2\mathfrak{Im}G^A_{A,B}(\omega)$ is the \textit{spectral density}. 
Invoking the fluctuation-dissipation theorem
$ G^<_{A,B}(\omega)= iA_{A,B}(\omega)/(1-e^{\beta\hbar\omega})$, the requirement of the equilibrium stability
$P\geq 0$ imposes
\( |\lambda_1|^2 \tilde{a} +|\lambda_2|^2 \tilde{a} +\lambda_1 \lambda_2^* \tilde{A} + \lambda_1^*\lambda_2 \tilde{A}^* \geq 0 \Longleftrightarrow \tilde{a} \geq  \frac{-\lambda_1 \lambda_2^* \tilde{A}+\text{c.c.}}{|\lambda_1|^2 +|\lambda_2|^2} \)
for all frequency $\omega$ and arbitrary values of $\lambda_1$ and $\lambda_2$. We therefore obtain
\( \tilde{a} \geq \text{Max}\left\{\frac{-\lambda_1 \lambda_2^* A_{12}+\text{c.c.}}{|\lambda_1|^2 +|\lambda_2|^2}\right\} =|\tilde{A}|.  \)

\subsection*{(vi) Concurrence as a measure of entanglement}
For a pure bipartite state $\rho_{\text{AB}}= \ket{\psi_{\text{AB}}}\bra{\psi_{\text{AB}}}$, we usually adopt the von Neumann entropy as the entanglement measure: $S(\ket{\psi_{\text{AB}}})\equiv -\tr\rho_A\ln \rho_A=-\tr\rho_B\ln \rho_B$.  For a general mixed state $\rho_{\text{AB}}$, this von-Neumann entropy is no longer a good measure since the classical mixture in $\rho_{\text{AB}}$ will have a nonzero contribution. We will adopt entanglement of formation as our entanglement measure.

The entanglement of formation is defined as
\( E_F(\rho_{\text{AB}}) \equiv \text{min} \sum_i p_i\, S(\ket{\psi^i_{\text{AB}}}),   \)
where the minimum is taken over all possible decompositions of $\rho_{\text{AB}}=\sum_i p_i \ket{\psi^i_{\text{AB}}}\bra{\psi^i_{\text{AB}}}$ and $S(\ket{\psi^i_{\text{AB}}})$ is the von Neumann entropy of the pure state $\ket{\psi^i_{\text{AB}}}$. Physically, $E_F(\rho_{\text{AB}})$ is the minimum amount of pure state entanglement needed to create the mixed state. This is extremely difficult to evaluate in general since we need to try all the decompositions. Quite remarkably an explicit expression of $E_F(\rho_{\text{AB}})$ is given when both $A$ and $B$ are two-state systems (qubits). This exact formula is based on the often used two-qubit concurrence, which is defined as \cite{PhysRevLett.80.2245}
\( \mathcal{C}(\rho)=\text{max} \{ 0,\lambda_1-\lambda_2-\lambda_3-\lambda_4 \}, \label{concurrence} \)
where $\lambda_i$'s are, in decreasing order, the square roots of the eigenvalues of the matrix $\rho (\sigma_y\otimes \sigma_y)\rho^*(\sigma_y\otimes \sigma_y)$, where $\rho^*$ is the complex conjugate of $\rho$. The entanglement of formation is then given by
\(   E_F(\rho)= h\Big( \frac{1+\sqrt{1-\mathcal{C}^2}}{2}  \Big), \;\text{with}\;\;    h(x)=-x\log_2x-(1-x)\log_2(1-x). \)
$E_F(\rho)$ is monotonically increasing and ranges from 0 to 1 as $\mathcal{C}(\rho)$ goes from 0 to 1, so that one can take the concurrence as a measure of entanglement in its own right. 
If our density matrix is in the form of
\( \rho=\mqty[\rho_{00} &  & & \rho_{03} \\  &\rho_{11}  &\rho_{12} &  \\  &\rho_{21}  &\rho_{22} &  \\  \rho_{30} &  & & \rho_{33} ],    \)
the expression of concurrence can be reduced to 
\( \mathcal{C}(\rho) =2 \max\{0, |\rho_{12}|-\sqrt{\rho_{00}\rho_{33}}, |\rho_{03}|-\sqrt{\rho_{11}\rho_{22}}   \}. \)
In  the situations corresponding to the plots in Fig.~\textcolor{red}{2} and Fig.~\textcolor{red}{3} in the main text, the concurrence is further reduced to $\mathcal{C}(\rho) =2|\rho_{12}|$.

\subsection*{(vii) Full entanglement dynamics}
In this section, we study the full entanglement dynamics governed by the master equation [Eq.~(\textcolor{red}{7}) in the main text]:
\(   \dv{}{t}\rho=-i[H_S,\rho]-\underline{\mathcal{L}}[\rho].   \)
Taking a trivial product state $\ket{\uparrow\downarrow}$ as the initial state, the time evolution of all the relevant elements of $\rho(t)$ is given by
\eq{\al{  \dot{\rho}_{00}&=-4 \tilde{a} \rho_{00}+2 a (\rho_{11}+\rho_{22})+2|A|(\rho_{12}+\rho_{21}),\\
 \dot{\rho}_{11}&=-2(a+\tilde{a})\rho_{11}+2 \tilde{a} \rho_{00}+2 a \rho_{33}-\Gamma(\rho_{12}+\rho_{21}),\\ 
  \dot{\rho}_{22}&=-2(a+\tilde{a} )\rho_{22}+2 \tilde{a} \rho_{00}+2 a \rho_{33}-\Gamma(\rho_{12}+\rho_{21}),\\
 \dot{\rho}_{33}&=-4 a \rho_{33}+2 \tilde{a} (\rho_{11}+\rho_{22})+2|\tilde{A}|(\rho_{12}+\rho_{21}),\\
 \dot{\rho}_{12}&=2\big[i\delta - (a+\tilde{a} ) \big]\rho_{12}-\Gamma(\rho_{11}+\rho_{22})+2|\tilde{A}|\rho_{00} +2|A|\rho_{33}\\ 
\dot{\rho}_{21}&=2\big[-i\delta - (a+\tilde{a}) \big]\rho_{21}-\Gamma(\rho_{11}+\rho_{22})+2|\tilde{A}|\rho_{00} +2|A|\rho_{33}, }\label{smmaster} }
with net dissipative coupling $\Gamma\equiv |A|+|\tilde{A}|$ and local field asymmetry $\delta\equiv (\Delta_1-\Delta_2)/2$ . In general, we need to solve the above six coupled differential equations with the constraint $\tr\rho(t)=1$.

\subsubsection{Derivation of Eq.~(8) in the main text}
At zero temperature,  $\tilde{a}=|\tilde{A}|=0$, the coupled differential equations above reduce to a single equation for $x\equiv \Re \bra{\uparrow\downarrow}\rho\ket{\downarrow\uparrow}$:
\( \ddot{x} + 4 a  \dot{x}+4 (\delta^2+ a^2-|A|^2)x=0, \)
with the  initial condition $\rho(t=0)=\ket{\uparrow\downarrow}\bra{\uparrow\downarrow}$.
We first note that, in our situation, $\rho_{33}(t)=0$ since $\rho_{33}(0)=0$ and the differential equation for $\rho_{33}$ is reduced to $\dot{\rho}_{33}=-4 a \rho_{33}$. We also have 
\( \dot{\rho}_{11}-\dot{\rho}_{22}=-2 a(\rho_{11}-\rho_{22})\longrightarrow \rho_{11}(t)=\rho_{22}(t) +e^{-2 a t}.  \)
Invoking $\tr\rho=1$, we obtain
 \( \rho_{00}(t) =1-2\rho_{22}(t) -e^{-2 a t}. \)
Choosing the free parameters to be $\rho_{22},\rho_{12}$ and $\rho_{21}$ we denote $\rho_{22}=h(t)$ and $\rho_{12}=x(t)+iy(t)$.
We can reduce Eq.~(\ref{smmaster}) to three coupled differential equations:
\eq{\al{\dot{h}&=-2 a h-2|A|x \\
   \dot{x}&=-2\delta y- 2 a x-2|A |h-|A|e^{-2 a t},\\
   \dot{y}&=2\delta x-2 a y, }\label{B3}}
and equivalently,
\eq{\al{ \dv{(he^{2at})}{t}&=-2|A|xe^{2a t}, \\
    \dv{(ye^{2at})}{t}&=2\delta x e^{2at},\\
     \dv{(xe^{2at})}{t}&=-2\delta (ye^{2at}) -2|A|(he^{2at})-|A|, }\label{B5} }
from which we obtain
\( \ddot{x} + 4a\dot{x}+4 (\delta^2+a^2-|A|^2)x=0.  \)

           \begin{figure}
  \includegraphics[scale=.32]{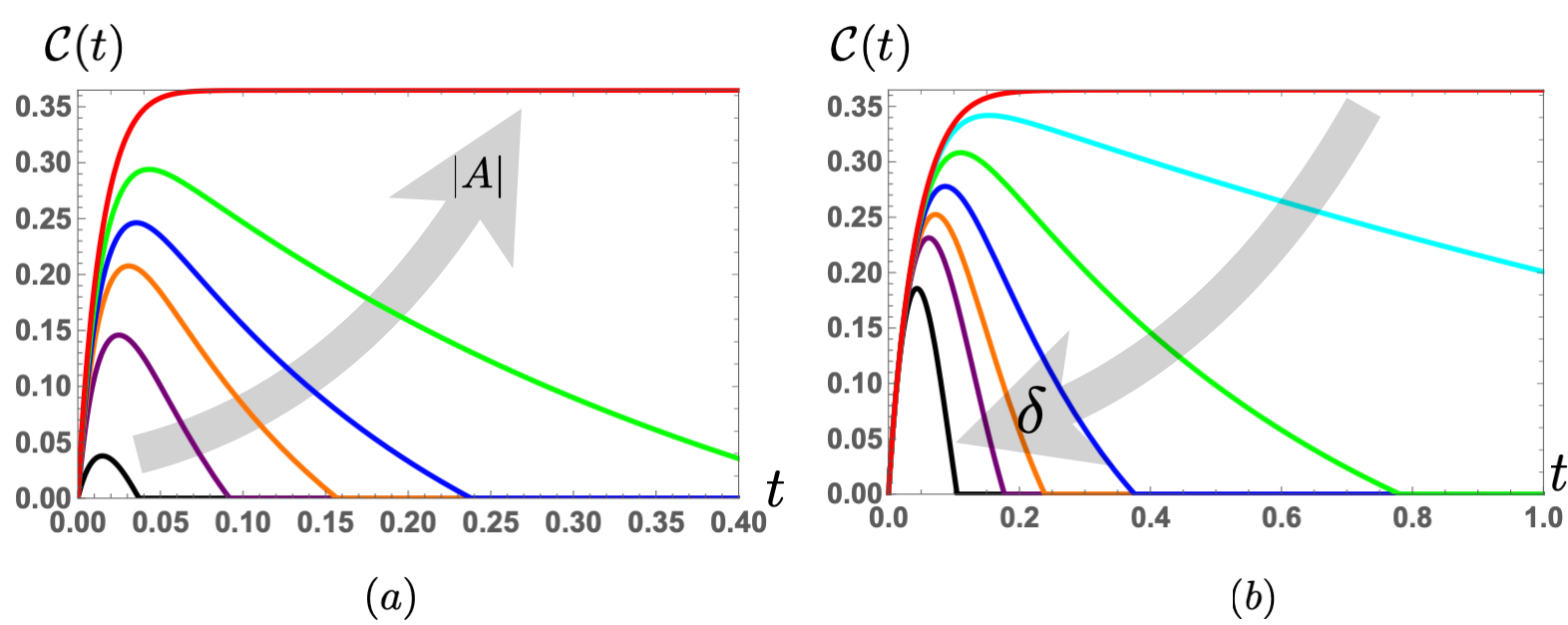}
   \caption{Concurrence as a function of time for the initial state $\ket{\uparrow\downarrow}$ at a finite temperature. (a). The local relaxations are set to $a=30$, $\tilde{a}=3$, and the local fields are equal $\delta=0$. Curves of different colors are plotted with an increasing dissipative coupling $|A|\rightarrow\{20,25,27,28,29,30\}$ along the direction of the gray arrow. The two dissipative couplings are related by $|A|/|\tilde{A}|=a/\tilde{a}=10$. When $|A|$ reaches its maximal values $30$ (and $|\tilde{A}|=|A|/10$ reaches its maximal allowed value $3$), we achieve steady entanglement (red curve). \,(b). The concurrence for $a=|A|=10, \tilde{a}=|\tilde{A}|=1$ and varying $\delta\rightarrow \{ 0, 2, 4, 6, 8, 10, 16     \}$. } 
   \label{smfig1}
\end{figure} 

Starting from this equation, one can solve for the density matrix $\rho(t)$ and thus the concurrence $\mathcal{C}(t)$ between the two qubits. For example, in the overdamped regime, 
  \( x(t)=-\frac{|A|}{2\kappa_0}\sinh2\kappa_0 t\, e^{-2at},\,\,\,y(t)=-\frac{|A|\delta e^{-2at}}{\kappa_0^2}\sinh^2\kappa_0 t,
  \,\,\, h(t)=\frac{|A|^2e^{-2at}}{\kappa_0^2}\sinh^2\kappa_0 t.   \)
    where $\kappa_0=\sqrt{|A|^2-\delta^2}$.  In this case, the full density matrix reads
   \(\rho(t)=\mqty[ 1- \frac{|A|^2\cosh2\kappa_0 t -\delta^2}{\kappa_0^2}e^{-2at}  & 0 & 0 & 0 \\  0& \frac{|A|^2\cosh^2\kappa_0 t-\delta^2}{\kappa_0^2}e^{-2at} & x(t)+iy(t) & 0 \\ 0& x(t)-iy(t) & \frac{|A|^2}{\kappa_0^2} e^{-2at}\sinh^2\kappa_0 t & 0 \\ 0 & 0 &0 & 0 ].\)
    The concurrence is thus given by 
       \( \mathcal{C}(t)=2|x(t)+y(t)|=2|A|\sinh\kappa_0 t e^{-2at} \sqrt{|A\cosh\kappa_0 t|^2 -\delta^2}/\kappa_0^2. \)
A time scale for the decay of the entanglement can be extracted:
   \( 1/\tau=2a-2\kappa_0=2\big(a-\sqrt{|A|^2-\delta^2} \big)\geq 0. \label{smlife}  \)
Therefore, to extend the lifetime of the entanglement, one can reduce the local field asymmetry $\delta$ or increase the dissipative coupling $|A|$. As one particular interesting scenario, equal local fields $\delta=0$ yields the lifetime of entanglement $\tau=1/(a-|A|)$, indicating that local relaxation $a$ and the dissipative coupling $|A|$ have perfectly opposite effects on entanglement.  
The maximal allowed value of $|A|=a$ can be reached when the spatial separation of  the two qubits are short enough. This length scale is set by the relevant excitations responsible for dissipation, such as magnons in a magnetic medium. When $|A|=a$, we can achieve a steady entanglement with concurrence $\mathcal{C}(\infty)=1/2$, with 
the final steady state being
\( \rho(\infty)=\frac{\ket{00}\bra{00}+  \ket{\uparrow\uparrow}\bra{\uparrow\uparrow} }{2},   \)
where $\ket{00}$ is the singlet state. It is also easy to versify that the state $\ket{00}$ and $\ket{\uparrow\uparrow}$ are the only two dark states in this situation. 
This can be seen from the effects of jump operators acting on the state: out of the four quantum-jump operators~\eqref{smjump}, only $J_3$ is operative and $J_3\ket{00}=0, J_3\ket{\uparrow\uparrow}=0$. Note that they are also dark states for the entire master equation if the induced effective Hamiltonian is XXZ, which is the form consistent with axial symmetry.   Combined with the fact that $J_3$ is invariant under the exchange of the two spins, it is clear that we can always achieve finite steady-state entanglement irrespective of the initial state as long as it is not totally symmetric.

   \begin{figure}
  \includegraphics[scale=.2]{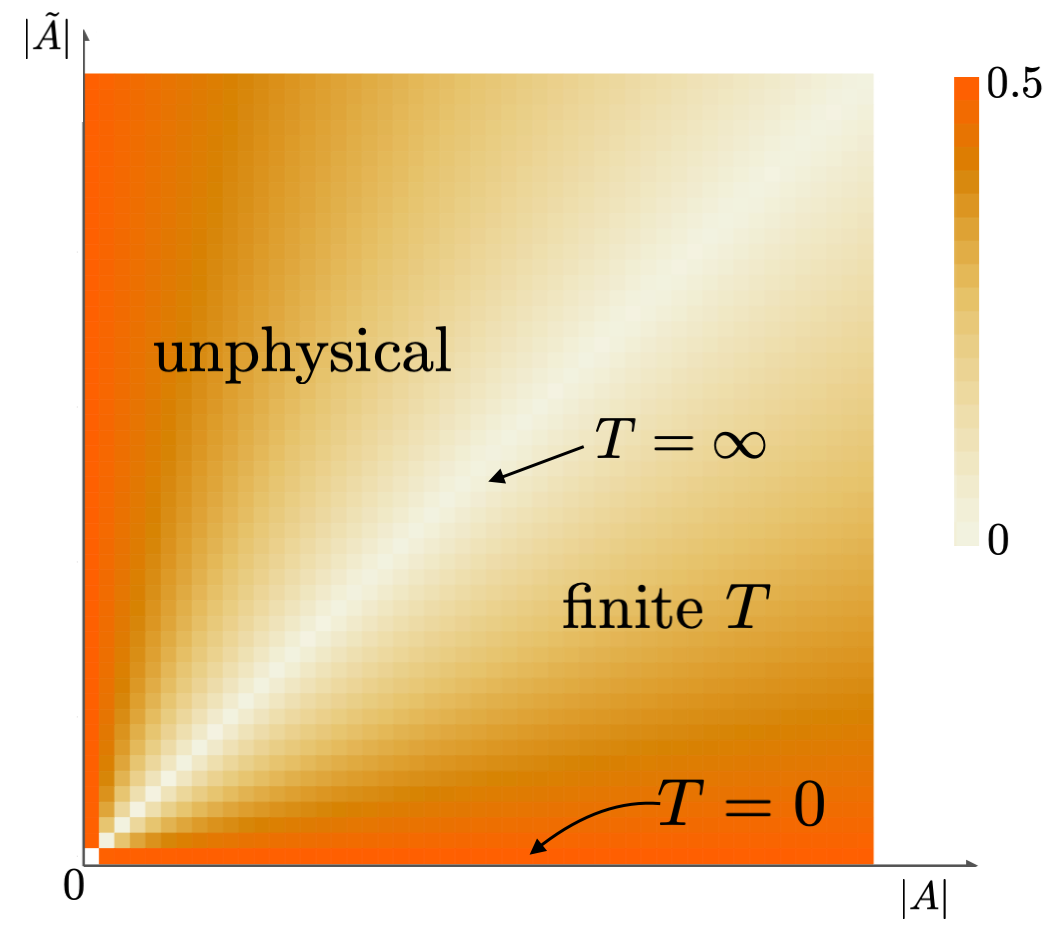}
   \caption{Final steady-state concurrence as a function of dissipative couplings $|A|$ and $|\tilde{A}|$, assuming they are at their maximal values $|A|=a$ and $|\tilde{A}|=\tilde{a}$, and $\delta = 0$.  The initial state is $\ket{\uparrow\downarrow}$.} 
   \label{smfig2}
\end{figure} 

\subsubsection{Dynamics of entanglement at finite temperature}
At finite temperatures, both local relaxations $a$, $\tilde{a}$ and dissipative couplings $|A|$, $|\tilde{A}|$ are nonvanishing. There is no analytic solution to the master equation \eqref{smmaster}. Instead, numerical solutions are studied. Similar to the conclusion we have drawn in the main text, we find that increasing the dissipative couplings $|A|$ and $|\tilde{A}|$ could extend the lifetime of entanglement dramatically and steady entanglement can be obtained when $|A|$ and $|\tilde{A}|$ reach their allowed maximal values $|A|=a$ and $|\tilde{A}|=\tilde{a}$. Recall that these four parameters are not independent: $a/\tilde{a}=|A|/|\tilde{A}|=e^{\beta \Delta}$, where $\beta= 1/k_B T$.
     
As shown in Fig.~\ref{smfig1} (a), with the local relaxations fixed $a=30$, $\tilde{a}=3$ and the local fields set to be equal $\delta=0$, the lifetime of entanglement increases as we increase the dissipative coupling $|A|$ (and also $|\tilde{A}|$ accordingly). A steady entanglement with the concurrence being around 0.35 is achieved when the dissipative couplings $|A|$ and $|\tilde{A}|$ reach their allowed maximal values $|A|=a=30$ and $|\tilde{A}|=\tilde{a}=3$.  In Fig.~\ref{smfig1} (b), we fix all dissipation parameters and vary $\delta$. The lifetime of entanglement decreases as $\delta$ increases, which is also consistent with the trend at zero temperature [see Eq.~\eqref{smlife}]. 

In Fig.~\ref{smfig2}, we show that a finite steady-state entanglement can always be achieved when the dissipative couplings reach their allowed maximal values $|A|=a$ and $|\tilde{A}|=\tilde{a}$, with $\delta = 0$. When $|A|=|\tilde{A}|$, the final steady entanglement is zero, corresponding to infinite temperature.  The entire $x$ axis with $|\tilde{A}|=0$ shows the  zero-temperature case, where the steady concurrence is $1/2$ as we have discussed before.  A steady entanglement smaller than $1/2$ persists for finite temperatures, in the regime $|\tilde{A}|/|A|=e^{-\beta \Delta}<1$, while $|\tilde{A}|>|A|$ is unphysical. At finite temperatures, the density matrix of the steady state is also partly made of the singlet state, which remains a dark state as $J_1\ket{00}=J_3\ket{00}=0$ (only these two jump operators are operative).


\begin{thebibliography}{93}
\bibitem [{iso()}]{isotropicspin}%
  \BibitemOpen
  \href@noop {} {}\bibinfo {note} {Here, the microscopic expression of $\mathcal{J}$ is
  $2\mathcal{J}=G^R_{S_1^x S_2^x}(0) + G^R_{S_2^x S_1^x}(0) =G^R_{S_1^y S_2^y}(0) +
  G^R_{S_2^y S_1^y}(0) =G^R_{S_1^z S_2^z}(0) + G^R_{S_2^z S_1^z}(0) $ when the
  spin space is isotropic.}\BibitemShut {Stop}%
\bibitem [{ent({\natexlab{b}})}]{entphase}%
  \BibitemOpen
  \href@noop {} {}\bibinfo {note} {We have gauged out the
  phases of $A, \tilde{A}$ by absorbing them into the definition of
  $\sigma^{\pm}$.}\BibitemShut {Stop}%
\bibitem [{gre()}]{greater}%
  \BibitemOpen
  \href@noop {} {}\bibinfo {note} {We use common convention for the greater
  Green's function: $G^>_{AB}(\omega)=-i\int^\infty_{-\infty}\langle
  A(t)B(0)\rangle e^{i\omega t}dt$.}\BibitemShut {Stop}%
\end{thebibliography}
\end{document}